\documentclass[12pt]{scrartcl}

\usepackage{amsmath, amssymb,amsthm,url}
\usepackage{graphicx}
\usepackage{subfig}
\newtheorem{lem}{Lemma}

\title{A single-item continuous double auction game}
\author{M. Ruijgrok\\
Mathematics Department, Utrecht University.\\
e-mail: m.ruijgrok@uu.nl}

\begin{document}

\maketitle
\begin{abstract}
A double auction game with an infinite number of buyers and sellers is introduced. All sellers posses one unit of a good, all buyers desire to buy one unit. Each seller and each buyer has a private valuation of the good. The distribution of the valuations define supply and demand functions. One unit of the good is auctioned. At successive, discrete time instances, a player is randomly selected to make a bid (buyer) or an ask (seller). When the maximum of the bids becomes larger than the minimum of the asks, a transaction occurs and the auction is closed. The players have to choose the value of their bid or ask before the auction starts and use this value when they are selected. Assuming that the supply and demand functions are known, expected profits as functions of the strategies are derived, as well as expected transaction prices. It is shown that for linear supply and demand functions, there exists at most one Bayesian Nash equilibrium. Competitive behaviour is not an equilibrium of the game. For linear supply and demand functions, the sum of the expected profit of the sellers and the buyers is the same for the Bayesian Nash equilibrium and the market where players behave competitively. Connections are made with the ZI-C traders model and the $k$-double auction. (JEL C720, D440) 
\end{abstract}
KEYWORDS:  Continuous double auction; supply and demand; Bayesian Nash equilibrium; ZI-C model; $k$-double auction.

\section{Introduction}
The continuous double auction (CDA) is an extensively studied institution, for two main reasons. First, many actual markets, such as commodity and stock exchanges, are of this type. Second, it provides a simple but effective model of a decentralized competitive market for one good.\\
Practically all studies on CDA's are either experimental or involve computer simulations. Theoretical results are lacking because of the complex nature of the CDA. In this paper, we consider an auction of a single item, where sellers and buyers all have some private valuation of this item. The auction participants are chosen successively and randomly to submit a bid (buyers) or an ask (sellers). Once the highest bid becomes equal or larger than the lowest ask, a transaction occurs and the auction is over. The profits for the participants involved in the transaction are equal to the difference between the transaction price and their private valuation.\\
Before the auction starts, each participant has to choose the value of the bid or ask he or she will make when selected, and this value cannot be changed during the auction. A buyer is therefore confronted with a strategic choice:  a high bid will increase his chance of winning the item, but decrease the profit he may gain, and similarly for a seller.\\
This paper will show that the payoff structure of this game is relatively simple. Moreover, for a large class of distributions of the private valuations, there exists a unique Bayesian Nash equilibrium, which can even be explicitly calculated. It will also be shown that the expected profits in this equilibrium have a remarkable property.\\
The study of the single-item CDA is intended as a first step in a theoretical analysis of the complete CDA, by which is meant an auction where, after the first item is auctioned, the successful buyer and seller leave the market and a new item is auctioned. This process then repeats until no more transactions are possible. Parsons (2006) quotes a number of authors who are pessimistic about the possibility of a theoretical analysis of the complete CDA. The results of this paper show that such pessimism is perhaps premature.\\
The distribution of the private valuations, which can be considered as maximal buying prices for the buyers and minimal selling prices for the sellers, define supply and demand functions and a corresponding competitive equilibrium price. Laboratory experiments, pioneered by Smith (1962), where complete CDA's were repeated a number of times, have shown that after a short learning period a market price emerges. Moreover, this price converges to the competitive equilibrium price, even though the subjects involved in the experiment were only aware of their own valuations. This finding is very robust and has been replicated many times, see Davis and Holt (1992), chapter 3.\\
It's not even necessary to be human to discover the equilibrium price in such repeated CDA's, as work inspired by the ZI-traders of Gode and Sunder (1993) has shown. Although Gode and Sunder were mainly interested in how the structure of a market influences its efficiency, their framework was later adopted by Cliff and Bruten (1997) to show that the price of the good in repeated CDA's populated by agents with a minimal learning capability generally converge to the competitive equilibrium price. These papers are at the origin of what has become a vast literature, of which Anufriev et. al. (2011) and Fano, LiCalzi and Pellizzari (2011) are recent examples.\\ 
One version of a complete CDA, namely where all bids and asks are cleared after one item has been auctioned, corresponds to a succession of single-unit continuous double auctions. Of course, the supply and demand functions change after every single item is auctioned, since the successful traders leave the auction. The game presented in this paper, already interesting in its own right, can therefore be seen as a building block of a game-theoretic model of a complete CDA. An evolutionary version of such a model could serve as a theoretical foundation for Smith's results and more generally for the emergence of competitive equilibrium in a decentralized market.\\
Most of the theoretical work on double auctions deals with call markets, see for instance Reny and Perry (2006). The type of model that comes closest to the one presented in this paper is that of the two-player $k$-double auction introduced by Chatterjee and Samuelson (1983). Although this auction involves only two participants, it can be reformulated as an auction with large, in fact infinite, sets of buyers and sellers. This set-up resembles our model of the single-item CDA, in particular in the form of the payoff functions.\\
To find a Bayesian Nash equilibrium (BNE) of either the $k$-double auction or the CDA, the expression for the expected payoff to a buyer or seller who adopts a certain strategy, given the strategies of all other auction participants, has to be derived. For the $k$-double auction this is much simpler than for the single-unit CDA. Satterthwaite and Williams (1989) show that the $k$-double auction generally has many BNE. In particular, it has a a two-parameter family of such equilibria. Using methods similar to the ones used by Satterthwaite and Williams, we will show that in contrast, the single-unit CDA has either no BNE, or a unique one. Moreover, for linear supply and demand functions, an exact expression for the BNE (if it exists) can be given for the single-unit CDA, whereas such an expression is only known for one special case of the $k$-double auction. Surprisingly, for this special case, when both maximum bidding prices and minimum selling prices are uniformly distributed on the interval $[0,1]$, these expressions are equal when $k=1/2$.\\
For the case of linear supply and demand functions, the expected profits for the buyers and sellers can be explicitly calculated. We compare these to the case that the market participants use a strategy corresponding to competitive equilibrium. This means that those who can afford it, bid or ask the competitive equilibrium price and others bid or ask what they want. We will show that this competitive strategy is not an equilibrium, but it still serves as a benchmark. We will show that the sum of the expected profits of the buyers and the sellers in the case of the BNE is equal to that in the case of the competitive strategy. However, the profit in the case of the BNE is spread out over a larger fraction of the population.\\
Finally, a bonus of this model is that it is possible to slightly adapt it so that it can be applied to the ZI-C model of Gode and Sunder. In this model, rather than choosing a unique value for their bid or ask, players select a random value constrained by their private valuation, when chosen. In particular, we derive an expression for the price distribution of the first transaction in the ZI-C model, a subject on which there has been some discussion, see Jahnsson (2011).\\\\
The paper is structured as follows. In section 2 the continuous double auction game will be defined and in section 3 the payoff functions are derived. In section 4 an expression for the distribution of transaction prices is derived and applied to the ZI-C model. In section 5 some general properties of Bayesian Nash equilibria are formulated. In section 6 an explicit expression for the BNE is given and a connection is made with the $k$-double auction. In section 7 it is shown that the competitive strategy is not an equilibrium. In section 8, the sum of the expected profits of buyers and sellers in the competitive strategy and the BNE are compared, again in the case of linear supply and demand functions. Section 9 concludes.

\section{The model}
In this auction, one unit of a certain good will be sold. The set of buyers is identified with the interval $[0,1]$, as is the set of sellers. 
Each buyer has a maximum price in the interval $[0,a]$, which he is willing to pay for the unit and each seller a minimum price in the interval $[0,b]$ for which she is willing to sell. Without loss of generality, we may assume $a=b=1$. These maximum and minimum prices define the type of the trader. \\
The assumption that the number of buyers and sellers and the number of types is (uncountably) infinite is meant as an approximation of the situation where these numbers are large but finite. It is straightforward to derive a finite model by replacing integrals in the payoff functions with finite sums. However, the results on the existence and the explicit expressions for the Bayesian Nash equilibria in the continuous model rely on solving differential equations, which is harder to translate to the finite case. Whether the results reported in this paper continue to hold for finite models will require a separate study. \\
The distributions of the seller and buyer types are common knowledge and follow from the supply and demand functions. For the buyers, we are given a strictly decreasing continuously differentiable demand function $D: [0,1]\rightarrow [d_-,d_+]$, with $0 \leq d_- < d_+ \leq 1$. If $p=D(q)$, then a fraction $q$ of the buyers is willing to pay a price $p$ or less for the unit. The supply side of the market is given by a strictly increasing and continuously differentiable supply function $S: [0,1]\rightarrow [s_-,s_+]$, with $0 \leq s_- < s_+ \leq 1$. If $p=S(q)$, then a fraction $q$ of the sellers is willing to sell for a price $p$ or more for the unit. We assume that $d_+>s_-$ and $s_+>d_-$, to guarantee that the graphs of the supply and demand functions intersect.\\
\begin{figure}[ht]
\centering
\subfloat[][]{\includegraphics[scale=0.35]{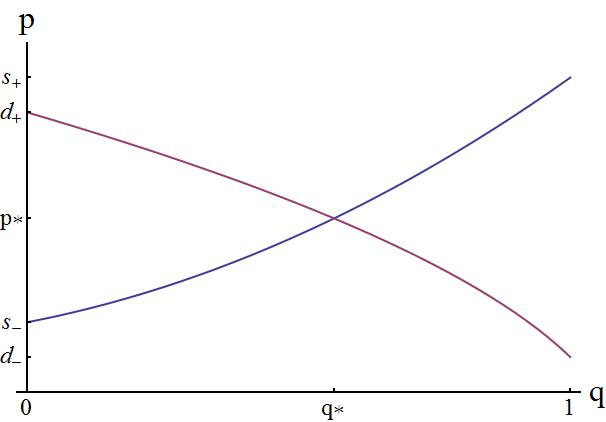}}
\qquad
\subfloat[][]{\includegraphics[scale=0.35]{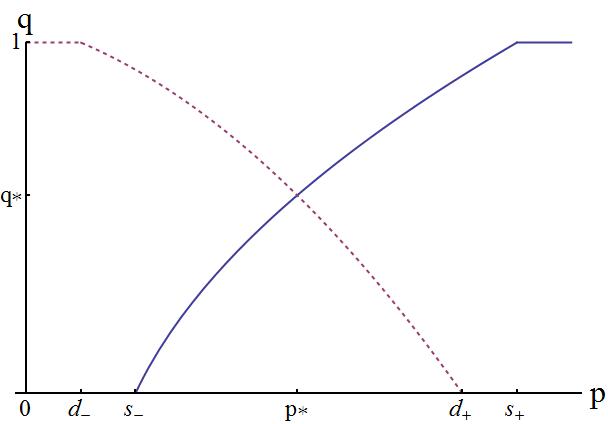}}
\caption{(a) The supply and demand functions.\\ (b) $S^{-1}(p)$ (solid line) and $D^{-1}(p)$ (dashed line).}
\label{vraagaanbod2}
\end{figure}
\noindent The inverses of the supply and demand function, $S^{-1}(m)$ and $D^{-1}(M)$, can be used to define the cumulative distribution functions of the types, see Figure \ref{vraagaanbod2}. If a seller is randomly chosen, then the probability that her type is less or equal to $m \in [s_-,s_+]$ is $S^{-1}(m)$. If $m<s_-$, this probability is zero and if $m>s_+$, it is equal to one. The probability density for the seller types is therefore given by 
\begin{align}
\sigma(m)=\begin{cases} \frac{d \, \, \, \, }{d m} S^{-1}(m), & \mbox{if } m\in (s_-,s_+) \\ 0, & \mbox{if } m\not\in (s_-,s_+) \end{cases} \, ,
\end{align}
Similarly, the probability that a randomly chosen buyer is of a type smaller than $M$, with $M \in [d_-,d_+]$  is $1-D^{-1}(M)$. If $M<d_-$ the probability is zero and if $M>d_+$ it is one. The probability density for the buyer types is therefore given by   
 \begin{align}
\mu(M)=\begin{cases} \frac{d \, \, \, \, }{d M}(-D^{-1}(M)), & \mbox{if } M\in (d_-,d_+) \\ 0, & \mbox{if } M\not\in (d_-,d_+) \end{cases} \, ,
\end{align}
At consecutive times $\tau=1$, $\tau=2$, $\ldots $, a trader is randomly and uniformly chosen. We can assume that the probabilities of choosing a buyer or a seller are equal. If, say, the fraction of buyers is larger than the fraction of sellers, then a certain amount of 'dummy' sellers with a minimum selling price of one can be added to even out the ratio of sellers to buyers.  When sellers outnumber buyers, extra 'dummy' buyers with maximum buying price zero can be introduced.\\ 
The chosen trader offers an ask (seller) or a bid (buyer). If an ask is offered which is less or equal to the maximum of the bids up to that time, a trade is effected between the seller who made the ask and the holder of the maximum bid, at a price equal to this maximum bid. In this case the seller is the {\it price-taker} and the buyer is the {\it price-maker}. Similarly, if a bid is placed which is larger or equal to the minimum of the asks, the trade is done between the buyer and the holder of the minimum bid, at a price equal to this minimum, with now the buyer as price-taker and the seller as price-maker. To replace the holder of the current minimum ask, a new ask must be strictly smaller. A similar rule applies to buyers.\\
The strategy of a buyer of type $M$ is the bid $x \in [0,1]$ he will offer, when given the opportunity. We assume that all traders with the same type, use the same strategy. The set of strategies of the buyers can therefore be described by a single function $b(M)$, known as the {\it strategy profile}, where $b: [d_-,d_+] \rightarrow [0,1]$. The strategies of the sellers are encapsulated by the strategy profile $a: [s_-,s_+] \rightarrow [0,1]$, so that a seller of type $m$ will ask $a(m)$.\\
We will assume that the minimum ask is strictly smaller than the maximum bid, so that the probability that the auction ends in finite time is equal to one.\\\\
We define the {\it outcome} of the transaction as the triple $\{ M,m,t\}$, where $M\in [d_-,d_+]$ and $m\in [s_-,s_+]$ are the types of the buyer and seller involved, and $t\in [0,1]$ the transaction price. From the outcome, we can calculate the profits of the traders involved, namely $M-t$ for the buyer and $t-m$ for the seller.\\
The outcome is a random variable, with a probability distribution that depends on the strategies used by the traders and on the distribution of the types of buyers and sellers. In particular we define 
\begin{align}
\label{betadef}
\beta(M,m,t;x,a(.),b(.))
\end{align}
as the probability density of the event that a buyer of type $M$ and a seller of type $m$ are involved in a transaction with transaction price $t$, given that the buyer uses strategy $x$ and the other buyers play $b(M)$ and the sellers $a(m)$.\\ 
Given $\beta(M,m,t;x,a(.),b(.))$, we define:
\begin{align}
\label{pib}
\pi_b(x,M;a(.),b(.))=\frac{1}{\mu(M)}\int_{s_-}^{s_+}\!\!\big [ \int_0^1\!\!(M-t)\beta(M,m,t;x,a(m),b(M))\,{\rm d}t\big ]\, {\rm d}m  \, .
\end{align}
The function $\pi_b(x,M;a(.),b(.))$, which we will often abbreviate as $\pi_b(x,M)$, is the probability density for the expected payoff of a buyer of type $M$ who plays strategy $x$, given that the rest of the buyers play $b(M)$ and the sellers $a(m)$. It means that the expected payoff to the group of buyers whose types are in some interval $S \subset [0,1]$ is given by
\begin{align*}
\int_S \mu(M)\pi_b(x,M) \, {\rm d}M\, ,
\end{align*}
assuming they all use the strategy $x$.
Similarly, 
\begin{align}
\label{alphadef}
\alpha(M,m,t;x,a(.),b(.))
\end{align}
is the probability density of the event that a buyer with limit $M$ and a seller with limit $m$ are involved in a transaction with transaction price $t$, given that the seller uses strategy $x$ and the other sellers play $a(m)$ and the sellers $b(M)$. With this expression we can define 
\begin{align}
\label{pia}
\pi_a(x,m;a(.),b(.))=\frac{1}{\sigma(m)}\int_{d_-}^{d_+}\!\!\big [ \int_0^1\!\!(t-m)\alpha(M,m,t;x,a(m),b(M))\,{\rm d}t\big ]\, {\rm d}M  \, .
\end{align}
The above expression is often abbreviated as $\pi_a(x,m)$.\\
It will be shown that $\pi_b(x,M)$ and $\pi_a(x,m)$ are continuous functions of $M$ and $m$, respectively. Therefore, the a-priori probability that a specific buyer or seller is involved in a transaction is zero, and so the expected profit of every trader is also zero. Nevertheless, we will take $\pi_b(x,M)$ and $\pi_a(x,m)$ as the payoff functions for types $M$ and $m$, respectively, and assume that the buyer of type $M$ will try to maximize $\pi_b(x,M)$ as a function of $x$, and similarly a seller of type $m$ will want to maximize $\pi_a(x,m)$.\\
This can be justified in the following way. Consider a model with a finite number of types $M_1, M_2, \ldots M_N$, and a finite number $K$ of buyers and let $f_i$ be the number of buyers of type $M_i$. Then we can approximate the profit of all the types $M_i=i/N$ by
\begin{align*}
\int_{M_i}^{M_i+1/N} \mu(M)\pi_b(x,M) \, {\rm d}M\, \approx \frac{\mu(M_i)}{N}\pi_b(x,M_i)\approx\frac{f_i}{K}\pi_b(x,M_i) \, .
\end{align*}
The profit for an individual buyer of type $M$ is then approximately $\pi_b(x,M_i)/K$ and it is rational for him to maximize this expression. Since we can consider the model as the limit $N, K \rightarrow \infty$ of a finite model, the only consistent choice for a payoff function for the buyers is $\pi_b(x,M)$. The payoff function for the sellers is taken to be $\pi_a(x,m)$.\\ 
We define a Bayesian Nash equilibrium of this game as a pair of strategy profiles $(a^*(m),b^*(M))$ such that $y=a^*(m)$ maximizes $\pi_a(y,m;a^*(.),b^*(.))$ for all $m\in [s_-,s_+]$ and $x=b^*(M)$ maximizes $\pi_b(x,M;a^*(.),b^*(.))$ for all $M\in [d_-,d_+]$. 

\section{Expected payoffs}
The ingredients that are still missing are expressions for $\pi_b(x,M)$ and $\pi_a(x,m)$. These will be constructed in a few steps.\\
We will take the standing maximum bid before $\tau=1$ to be zero, and the minimum ask at that time to be one and also note that the probability that a transaction occurs at $\tau=1$ is equal to zero.\\
The event that a buyer of type $M$ and a seller of type $m$ are involved in a transaction with transaction price $t$, given that the buyer uses strategy $x$, can be decomposed in a countably infinite set of mutually disjoint events, namely the event that the transaction occurs at $\tau=2$, at $\tau=3$, etc. If there was a transaction at $\tau=n$, then there was no transaction at $\tau=n-1$ and one of two things happened at $\tau=n$. Either an ask was made that was lower or equal to the highest standing bid, or a bid was made that was higher or equal to the lowest standing ask. It follows that we can write, with some abuse of notation:
\begin{align}
\label{betaexpr}
&\beta(M,m,t;x,a(.),b(.))=\nonumber \\
&\sum\limits_{n=2}^\infty Pr({\rm no\: transaction\: at\:} \tau=n-1 \:{\rm \:and\: holder\: maximum\: bid\: is\:of\:type\:} M)\nonumber\\
&Pr({\rm seller\: of\:type\:} m {\rm \:is \:chosen}) Pr(a(m)\leq x)\, Pr(t=x)\, + \nonumber \\
&\sum\limits_{n=2}^\infty Pr({\rm no\: transaction\: at\:} \tau=n-1 {\rm \:and\: holder\: minimum\: ask\: is\:of\:type\:}m)\nonumber\\ 
&Pr({\rm buyer\: of\:type\:} M {\rm \:is \:chosen})\, Pr(a(m)\leq x)\, Pr(t=a(m)).
\end{align}
Note that the events $(a(m)\leq x)$, $(t=x)$ and $(t=a(m))$ are not chance events, so their probabilities are degenerate.\\
In Appendix \ref{derivout} it is shown that a closed expression exists for (\ref{betaexpr}), in terms of the cumulative distribution function of the asks:
\begin{align}
\label{defA}
A(x)=Pr({\rm ask}\leq x)= \int_{s_-}^{s_+}\!\!\theta(x-a(m))\,\sigma(m)\, {\rm d}m \, ,
\end{align}  
and the bids
\begin{align}
\label{defB}
B(x)=Pr({\rm bid}\leq x)= \int_{d_-}^{d_+}\!\!\theta(x-b(M))\,\mu(M)\, {\rm d}M \, ,
\end{align}
where $\theta(z)$ is defined as $\theta(z)=0$ if $z<0$ and $\theta(z)=1$ if $z \geq 0$.\\
We also define 
\begin{align}
\label{defcalAB}
{\cal A}(x)&=Pr({\rm ask}< x) \nonumber\\
{\cal B}(x)&=Pr({\rm bid}< x) \, .
\end{align}
Then
\begin{align}
\label{alphatotaal}
&\alpha(M,m,t;x,a(m),b(M))\,= \nonumber \\
&\mu(M)\sigma(m)\theta(b(M)-x)\left(\gamma_2(x)\delta(t-x)+\gamma_1(b(M))\delta(t-b(M))\right)\, ,
\end{align}
and
\begin{align}
\label{betatotaal}
&\beta(M,m,t;x,a(m),b(M))\,= \nonumber \\
&\mu(M)\sigma(m)\theta(x-a(m))\left(\gamma_1(x)\delta(t-x)+\gamma_2(a(m))\delta(t-a(m))\right)\, .
\end{align}
Here, the functions $\gamma_1(x)$ and $\gamma_2(x)$ are defined as:
\begin{align}
\label{gamma12}
\gamma_1(x)=\frac{1}{(1-B(x)+A(x))(1-{\cal B}(x)+A(x))} \, ,
\end{align}
and
\begin{align}
\label{gamma22}
&\gamma_2(x)=\frac{1}{(1-{\cal B}(x)+A(x))(1-{\cal B}(x)+{\cal A}(x))}.
\end{align}
Momentarily disregarding the difference between probability and probability density, we can interpret (\ref{betatotaal}) as follows. The probability that a buyer of type $M$, playing strategy $x$, is involved in a transaction with a seller of type $m$, playing strategy $a(m)$ is proportional to the probability that a buyer of type $M$ and a seller of type $m$ are chosen in the process, explaining the term $\mu(M)\sigma(m)$. Also, a necessary condition for a transaction to occur between this buyer and this seller is that $x\geq a(m)$, which accounts for the term $\theta(x-a(m))$. Finally, the transaction price is either $t=x$, which happens with a probability $\gamma_1(x)$, or $t=a(m)$, which happens with a probability $\gamma_2(a(m))$, thus leading to the last term. The surprising aspect of (\ref{alphatotaal}) and (\ref{betatotaal}) is that the expressions for $\gamma_1(x)$ and $\gamma_2(x)$ are so simple. 
\\\\
To derive an expression for the expected payoff $\pi_b(x,M;a(m),b(M))$, we substitute (\ref{betatotaal}) in (\ref{pib}). This yields
\begin{align}
\label{intoverm}
&\pi_b(x,M;a(m),b(M))= \nonumber \\
&\int_{s_-}^{s_+}  \int_0^1 (M-t)\sigma(m)\theta(x-a(m))\gamma_1(x)\delta(t-x)\, {\rm d}t\, {\rm d}m \, + \nonumber \\ 
&\int_{s_-}^{s_+}  \int_0^1 (M-t)\sigma(m)\theta(x-a(m))\gamma_2(a(m))\delta(t-a(m))\, {\rm d}t\, {\rm d}m \, .
\end{align}
The first part of (\ref{intoverm}) is easily integrated:
\begin{align}
\label{easybit}
&\int_{s_-}^{s_+}  \int_0^1 (M-t) \sigma(m)\theta(x-a(m))\gamma_1(x)\delta(t-x)\, {\rm d}t\, {\rm d}m\, = \nonumber \\
&A(x)\gamma_1(x)(M-x)\, .
\end{align}
For the second part of (\ref{intoverm}), we use the following
\begin{lem} \label{lem1} Let
\begin{enumerate}
\item $f(z) : [0,1] \rightarrow {\mathbb R}$ be a measurable function.
\item $g(t) : [a,b] \rightarrow {\mathbb R}$ be a continuous function.
\item $\sigma(m) : [0,1] \rightarrow {\mathbb R}$ be a continuous probability density.
\item $a(m) : [s_-,s_+] \rightarrow [0,1]$ and $b(M) : [d_-,d_+] \rightarrow [0,1]$ be increasing, a.e. continuously differentiable strategy profiles. Denote $a(s_{\pm})=a_{\pm}$ and $b(d_{\pm})=b_{\pm}$
\end{enumerate}
Then 
\begin{align}
&\int_{s_-}^{s_+} \int_0^1 g(t)\sigma(m)f(a(m))\delta(t-a(m))\, {\rm d}t\, {\rm d}m=\int_0^1 g(q)f(q)\,{\rm d}A(q) \label{lemeq1}\\
&and \nonumber\\
&\int_{d_-}^{d_+} \int_0^1 g(t)\mu(M)f(b(m))\delta(t-b(M))\, {\rm d}t\, {\rm d}M=\int_0^1 g(q)f(q)\,{\rm d}{\cal B}(q). \label{lemeq2}
\end{align}
\end{lem}
\noindent
{\bf Proof}: See Appendix \ref{proof}\\\\
Applying equation (\ref{lemeq1}) to $f(z)=\theta(x-z)\gamma_2(z)$ and $g(t)=M-t$, we find:
\begin{align}
\label{diffbit}
& \int_{s_-}^{s_+}\int_0^1(M-t)\sigma(m)\theta(x-a(m))\gamma_2(a(m))\delta(t-a(m))\, {\rm d}t\, {\rm d}m \, =\nonumber \\
&\int_0^1 (M-q)\theta(x-q)\gamma_2(q)\, {\rm d}A(q) \, .
\end{align}
Substituting (\ref{easybit}) and (\ref{diffbit}) in (\ref{intoverm}), we finally find the expression for the expected profit of a buyer of type $M$ who plays $x$, while other buyers play $b(M)$ and the sellers $a(m)$:
\begin{align}
\label{pib2}
\pi_b(x,M;a(.),b(.))=\left((M-x)A(x)\gamma_1(x)+\int_0^x\!\!(M-q)\gamma_2(q)\, {\rm d}A(q)\right) \, .
\end{align}
A similar expression for $\pi_a(x,m;a(.),b(.))$ can be found:
\begin{align}
\label{pia2}
\pi_a(x,m;a(.),b(.))=\left((x-m)(1-{\cal B}(x))\gamma_2(x)+\int_x^1\!\!\!\!(q-m)\gamma_1(q)\, {\rm d}{\cal B}(q)\right) \, .
\end{align}
We note that the expected profit of a buyer has two components. The first one, involving a term $M-x$ corresponds to transactions in which the buyer is the price maker. The second one is an integral over $q$, involving the term $M-q$. This term comes from transactions where the buyer is the price taker. 
\section{Distribution of transaction prices and connection with ZI-C traders}
\begin{figure}[tb]
\centering
\includegraphics[scale=0.4]{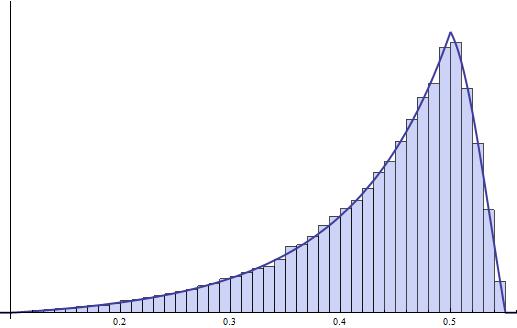}
\caption{Distribution of transaction prices for the case $D(x)=0.55-0.05x$ and $S(x)=0.1+0.7x$. The histogram shows the result of a numerical simulation with $50000$ transactions, the solid line is the plot of the right-hand side of (\ref{Tdistr}). The average transaction price is $\bar{T}=0.438$, while the competitive price is $p^*=0.52$.}
\label{ZI}
\end{figure}

Apart from giving expressions for expected profits, equations (\ref{alphatotaal}) and (\ref{betatotaal}) can also be used to derive the probability distribution of the transaction prices, in case the strategy profiles are continuous, differentiable and strictly increasing. Then $A(t)={\cal A}(t)$ and $B(t)={\cal B}(t)$ are also continuous, differentiable and strictly increasing on $[s_-,s_+]$ and $[d_-,d_+]$, respectively. Also, $\gamma_1(t)=\gamma_2(t)$$=(1-B(t)+A(t))^{-2}$ for all $t \in [0,1]$.\\
Let the strategy profiles $a(m)$ and $b(M)$, and hence $A(x)$ and $B(x)$, be given and let $T(t)=Pr(transactionprice\leq t)$ be the cumulative distribution function of the transaction prices. Then it is shown in Appendix \ref{trans} that
\begin{align}
\label{Tdistr}
T(t)=\frac{A(t)}{(1-B(t)+A(t))}\, .
\end{align}
The above result is also relevant for the Zero Intelligence Constrained model of Gode and Sunder (1993). The set-up of this model is similar to that of the game defined in this paper. However, the ZI-C players play a random strategy, where buyers offer a price which is randomly and uniformly chosen between zero and the players type and sellers randomly and uniformly choose a bid between their type and one. Another difference is that there is only a finite number of traders, who leave the market after a successful transaction.\\
Since the result (\ref{Tdistr}) and its derivation only depend on $A(x)$ and $B(x)$, it is clear that it can be applied to the ZI-C model, when there is an infinite number of traders and we only consider the first transaction.\\
Gode and Sunder claim on the basis of simulations that the transaction price converges during one trading session to the equilibrium price, but Cliff and Bruten (1997) contend that this happens only for special supply and demand functions. Jahnson (2011) provides an extensive analysis of this debate, which favours Gode and Sunder.\\
Also, Othman (2008) shows that the analytical foundation of Cliff and Bruten's critique is incorrect. Based on a large number of simulations, he finds that the distribution of first transaction prices is given by (\ref{Tdistr}), but does not provide a mathematical justification.\\
It remains true, however, that this distribution can have an average which is not close to the equilibrium price, as the example shown in figure \ref{ZI} demonstrates.
\section{Bayesian Nash equilibria}
\begin{figure}
\centering
\subfloat[][]{\includegraphics[scale=0.4]{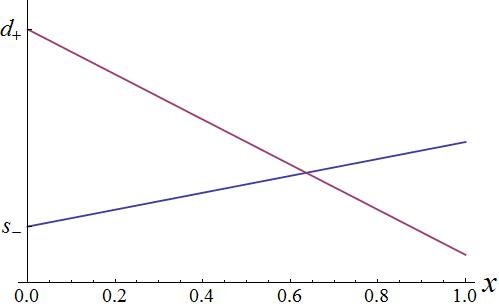}}
\qquad
\subfloat[][]{\includegraphics[scale=0.4]{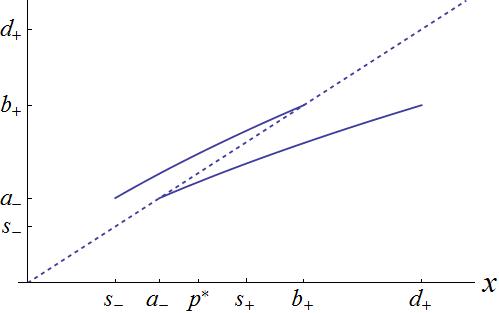}}
\caption{(a) Linear supply and demand functions $S(x)$ and $D(x)$. \\ (b) The functions $a(x)$ and $b(x)$ corresponding to the BNE.}
\label{supdem}
\end{figure}

Let us first note that associated with every Bayesian Nash equilibrium, henceforth BNE, there exist in general many other BNE that are essentially equivalent to it. For every pair of strategy profiles $(a(m),b(M))$, we can define the {\it extramarginal} sellers as those whose ask is higher than the maximum bid. Similarly, we can define extramarginal buyers as those whose bid is smaller than the minimum ask. These extramarginal traders will never be involved in a trade. Consequently, we will focus on the strategies of the other traders, who will be referred to as {\it intramarginal} traders.\\ 
In this section we will limit ourselves to continuous, differentiable and strictly increasing strategy profiles. For ease of notation, introduce
\begin{align*}
B_c(x)=1-B(x)=Pr(bid \geq x) \, .
\end{align*}
We have
\begin{align}
\label{defA2}
A(x)= \begin{cases} 0 & \mbox{if } 0\leq x \leq a_-  \\ \int_{s_-}^{a^{-1}(x)}\!\!\sigma(m)\, {\rm d}m=S^{-1}(a^{-1}(x)), & \mbox{if } a_-\leq x \leq a_+ \\ 1 & \mbox{if } a_+\leq x \leq 1 \, , \end{cases}
\end{align}
and
\begin{align}
\label{defB2}
B_c(x)= \begin{cases} 1 & \mbox{if } 0\leq x \leq b_- \\ \int_{b^{-1}(x)}^{d_+}\!\,\mu(M)\, {\rm d}M=D^{-1}(b^{-1}(x)), & \mbox{if } b_-\leq x \leq b_+ \\ 0 & \mbox{if } b_+\leq x \leq 1 \, , \end{cases}
\end{align}
where $S(x)$ and $D(x)$ are the supply and demand function, respectively. It follows immediately from the above and the assumptions on $a(m)$, $b(M)$, $S(x)$ and $D(x)$ that $A(x)$ and $B_c(x)$ are strictly increasing and strictly decreasing, respectively and both are continuously differentiable, except in $a_{\pm}$ and $b_{\pm}$, respectively.\\
The functions $\pi_a(x;m)$ and $\pi_b(x;M)$ are therefore differentiable with respect to $x$, except in $a_{\pm}$ and $b_{\pm}$ where jumps in the derivatives may occur. We find that the derivatives of $\pi_a(x;m)$ and $\pi_b(x;M)$, where they exist, are given by:
\begin{align}
\label{forder}
&\frac{\!\!{\rm d}}{{\rm d}x}\pi_a(x;m)= \nonumber \\
&\big [ 2(m-x)(-B'_c(x)A(x)+B_c(x)A'(x))+B_c(x)(B_c(x)+A(x)) \big ](B_c(x)+A(x))^{-3}  \nonumber \\
&\frac{\!\!{\rm d}}{{\rm d}x}\pi_b(x;M)= \nonumber \\
&\big[2(M-x)(-B'_c(x)A(x)+B_c(x)A'(x))-A(x)(B_c(x)+A(x))\big ](B_c(x)+A(x))^{-3} \nonumber \\
\end{align} 
Some properties of the equilibrium solutions are expressed in the following\\
\begin{lem} 
\label{lem2} 
Let $a :[s_-,s_+]\rightarrow [0,1]$ and $b :[d_-,d_+]\rightarrow [0,1]$ be continuous and strictly increasing and let $y=a(m)$ maximize $\pi_a(y,m;b(.),a(.))$ for all $m\in [s_-,s_+]$ and $x=b(M)$ maximize $\pi_b(x,M;b(.),a(.))$ for all $M\in [d_-,d_+]$. Denote $a(s_{\pm})=a_{\pm}$ and $b(d_{\pm})=b_{\pm}$. Then
\begin{enumerate}
\item $a(m)> m$, for all $m\in [s_-,b_+)$ and $b(M)< M$, for all $M\in (a_-,d_+]$.
\item $a_- \geq d_-$ and $b_+\leq s_+$.
\item $a(b_+)=b_+$ and $b(a_-)=a_-$.
\item The set of intramarginal sellers is $[s_-,b_+]$ and the set of intramarginal buyers is $[a_-,d_+]$.
\end{enumerate}
\end{lem}
\noindent{\bf Proof}:
\begin{enumerate}
\item From (\ref{forder}) it follows that $\frac{\!\!{\rm d}}{{\rm d}x}\pi_a(x;m)|_{x=m}=B_c(x)(B_c(x)+A(x))> 0$, for $m\in [s_-,b_+)$. Therefore, $\pi_a(y,m)$ is maximized in a value $y=a(m)> m$. The proof that $b(M)< M$ on $(a_-,d_+]$ is similar. 
\item Assume $a_-<d_-$ and $a_-<b_-$. This implies that the ask of the seller of type $s_-$ is lower than the smallest possible bid, so this seller is a guaranteed winner, if she is chosen. It is clear that this seller can increase her expected profit by choosing a strategy $a'$, with $a_-<a'<b_-$, since she will remain a sure winner, but this time with a higher profit when she is the price-maker and the same profit when she is the price-taker. Therefore, $a(m)$ does not maximize the expected profit of the sellers.\\
Assume $a_-<d_-$ and $a_-\geq b_-$. The expected profit for a buyer of type $d_-$ is then zero. He can improve this to a positive value by choosing a strategy $b'$ with $a_-<b'<d_-$, because then he has a positive probability of being involved in a transaction, when chosen, with a positive profit for him. Thus, $b(M)$ does not maximize the expected profit of the buyers. From this contradiction we conclude that $a_- \geq d_-$. The proof that $b_+\leq s_+$ is similar.
\item We have that $\pi_b(x,M)=0$ for $x\leq a_-$. Also, because $A(x)>0$ for $x>a_-$, we find from (\ref{pib2}) that $\pi_b(x,M)>0$ for $a_-<x<M$. Combining with part 1 of this lemma, we conclude that $a_-<b(M)<M$, for $a_-<M<d_+$. Taking $M\downarrow a_-$ and using continuity of $b(M)$, it follows that $b(a_-)=a_-$. The proof that $a(b_+)=b_+$ is similar. 
\item Let $m>b_+$. Since $a(m)$ is strictly increasing, $a(m)>a(b_+)=b_+$, therefore no seller of type larger than $b_+$ will ever participate in a transaction. A similar argument shows that the intramarginal buyers constitute the set $[a_-,d_+]$. 
\end{enumerate}
A graphical representation of these properties is given in Figure \ref{supdem}.

\subsection{First order conditions}
\begin{figure}
\centering
\subfloat[][]{\includegraphics[scale=0.5]{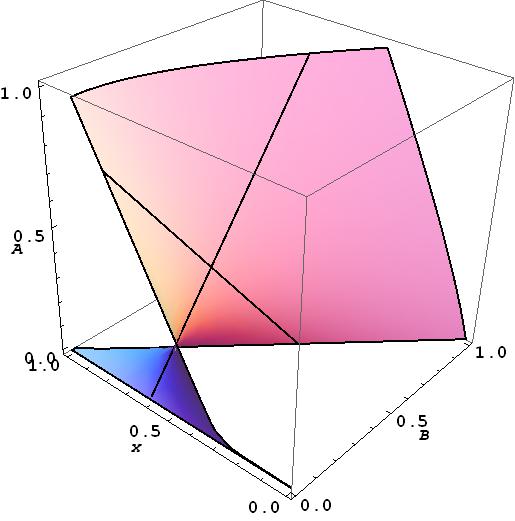}}
\qquad
\subfloat[][]{\includegraphics[scale=0.5]{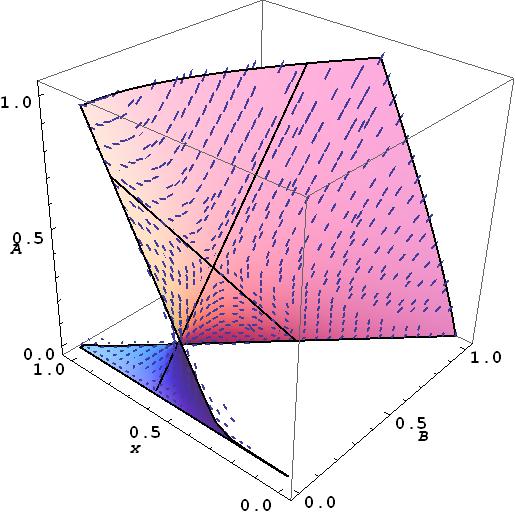}}
\caption{The surface $V$ and the vector field (\ref{vecfield}) on it, for the case $S(x)=0.3+0.6x$ and $D(x)=1-x$. Here, $s_+=0.9$, $d_-=0$, $a_-=0.43$ and $b_+=0.78$, so that the seperatrix connecting the plane $A=0$ with $B=0$ constitutes the unique BNE.}
\label{surfaces}
\end{figure}

To systematically derive BNE of the game, we first note that in (\ref{forder}) we can replace the parameter $m$ by a function of $x$, on the interval $[s_-,b_+]$, by using the fact that $x=a(m)$ on this interval. Therefore, $m=a^{-1}(x)=S(A(x))$. Similarly, $M=D(B_c(x))$ on $[a_-,d_+]$.\\
The first order conditions for a maximum then become
\begin{align*}
\frac{\!\!{\rm d}}{{\rm d}x}\pi_a(x;m)=0 \quad , \quad \frac{\!\!{\rm d}}{{\rm d}x}\pi_b(x;M)=0 \, .
\end{align*}
However, this set of equations is not independent, as is easily seen from (\ref{forder}). Rather, we have one differential equation and one consistency equation:
\begin{align}
&2(S(A(x))-x)(-B'_c(x)A(x)+B_c(x)A'(x))+B_c(x)(B_c(x)+A(x))=0 \, , \label{difeqn} \\
&B_c(x)(D(B_c(x))-x)=A(x)(x-S(A(x)))\, , \, {\rm for}\,\, x\in [a_-,b_+]\, . \label{consist}
\end{align}
There are also the following boundary conditions :
\begin{align*}
A(a_-)=0\, , \quad B_c(b_+)=0\, . 
\end{align*}
The values of $a_-$ and $b_+$ are not yet determined. Note that from the boundary conditions and the consistency equation (\ref{consist}), we have 
\begin{align*}
A(b_+)=S^{-1}(b_+)\, , \quad B_c(a_-)=D^{-1}(a_-) \, ,
\end{align*}
which also follows from Lemma 2.3 and (\ref{defA2}) and (\ref{defB2}). Also, $A(x)$ must be strictly increasing and $B_c(x)$ strictly decreasing.\\
A practical way to find the solutions $A(x)$ and $B_c(x)$ and the values of $a_-$ and $b_+$ is to use (\ref{consist}) to express $A(x)$ in terms of $B_c(x)$ (or the other way round) and substitute in (\ref{difeqn}). Together with the boundary condition $A(a_-)=0$, this then yields a solution $A(x;a_-)$, with $a_-$ a free parameter. This expression can in turn be used to construct $B_c(x;a_-)$. Solving $B_c(x;a_-)=0$ then gives $b_+(a_-)$. It would seem that this procedure yields a one-parameter family of solutions, parametrized by $a_-$. We will show in the next section that at least for linear supply and demand functions this is not true: in that case the BNE either does not exist or it is unique.\\
As a preparation, we consider the system (\ref{difeqn}), (\ref{consist}) from a geometric viewpoint. We introduce a time-like parameter $\tau$ and construct a vector field ${\bf F}(A,B_c,x)$ on $[0,1]^3$, such that the solution curves of the corresponding flow, defined by 
\begin{align*}
\frac{d}{d \tau}(A(\tau),B_c(\tau),x(\tau))^t={\bf F}(A,B_c,x)
\end{align*}
are solutions of (\ref{difeqn}), (\ref{consist}).\\
We first note that (\ref{difeqn}) can be interpreted as stating that a certain 1-form vanishes for all $(A,B_c,x)$:
\begin{align}
\label{1form}
2(S(A)-x)B_c\, dA-2(S(A)-x)A \,dB_c +B_c(B_c+A)\,dx=0\, .
\end{align}
The solution curve of the flow through a point $(A, B_c, x)$ has tangent vector $(dA, dB_c, dx)$, which according to (\ref{1form}) lies in a plane tangent to:
\begin{align*}
{\bf v_1}=(2(S(A)-x)B_c,-2(S(A)-x)A,B_c(B_c+A))^t\, .
\end{align*}
Secondly, the consistency condition (\ref{consist}) defines a surface $V$ in $(A,B_c,x)$ space, to which the solutions are confined. If we write (\ref{consist}) in the form $f(A,B_c,x)=0$, then it follows that the tangent vector of a solution curve through $(A,B_c,x)$ must lie in a plane tangent to
\begin{align*}
{\bf v_2}=\nabla f(A,B_c,x) \, .
\end{align*}
It then follows that the vector field given by 
\begin{align}
\label{vecfield}
{\bf F}(A,B_c,x)={\bf v_1}\times {\bf v_2}
\end{align} 
yields solution curves that are solutions to (\ref{difeqn}), (\ref{consist}), provided the initial condition lies on the plane $V$. Figure 5 shows the surface $V$ and the vector field (\ref{vecfield}) on it, in the case $S(x)=0.3+0.6x$ and $D(x)=1-x$.
As will be shown in the next section, for linear $S(x)$ and $D(x)$ the vector field has a singular point on $V$, of saddle type. This saddle point has a stable and an unstable manifold, the separatrices, which are also are shown in Figure \ref{surfaces}. From this figure it is clear that the separatrix connecting the plane $B_c=0$ with the plane $A=0$ is the only possible solution curve which fulfils the boundary conditions.\\
However, it might happen that the separatrix intersects the plane $B_c=0$ in a value $x>s_+$. This would imply that $b_+=x>s_+$, which contradicts Lemma 2.2. In such a case, there does not exist a BNE. Similarly if the intersection of the separatrix with the plane $A=0$ is in a $x<d_-$, the BNE does not exist. In other cases, (\ref{difeqn}), (\ref{consist}) has a solution and it is unique.
\section{Special case: $S(x)$ and $D(x)$ linear functions}
We now consider the special case that the supply function is given by $S(x)=s_-+\alpha x$ and the demand function by $D(x)=d_+-\beta x$, for $x \in [0,1]$ and construct explicit expressions for $A(x)$ and $B_c(x)$. This corresponds to the situation that the types of the sellers are uniformly distributed over the interval $[s_-,s_+]$ and the types of the buyers uniformly over $[d_-,d_+]$.\\
The system (\ref{difeqn}), (\ref{consist}) becomes:
\begin{align}
&2(d_+-\beta B_c-x)(-B_c'A+A'B_c)-A(A+B_c)=0 \label{eqlin}\\
&B_c(d_+-\beta B_c-x)=A(x-s_--\alpha A)\, \label{condB}. 
\end{align}
It is shown in appendix \ref{solution} that an explicit solution of (\ref{eqlin}), (\ref{condB}) is given by:
\begin{align}\label{Aexpr}
&A(x)= \nonumber\\
&\frac{-\gamma \alpha (d_+-x)+(2-\gamma)\beta (x-s_-)}{2\beta \alpha (d_+-s_-)}(d_+-s_-+\frac{\alpha(d_+-x)+\beta(x-s_-)}{2\sqrt{\alpha \beta}})\, , 
\end{align}
and
\begin{align}\label{Bexpr}
&B_c(x)= \nonumber\\
&\frac{(1+\gamma)\alpha (d_+-x)-(1-\gamma)\beta (x-s_-)}{2\beta \alpha (d_+-s_-)}(d_+-s_-+\frac{\alpha(d_+-x)+\beta(x-s_-)}{2\sqrt{\alpha \beta}})\, , 
\end{align}
with
\begin{align*}
\gamma=\frac{\sqrt{\beta}}{\sqrt{\alpha}+\sqrt{\beta}}\, .
\end{align*}
To find $a_-$, we solve $A(x)=0$ by equating the first factor in (\ref{Aexpr}) to zero and using that $\gamma^2\alpha-(1-\gamma)^2\beta=0$. This gives
\begin{align}
\label{amin}
a_-=s_-+(1-\gamma)^2(d_+-s_-) \, .
\end{align} 
A similar procedure shows that
\begin{align}
\label{bplus}
b_+=d_+-\gamma^2(d_+-s_-)\, .
\end{align}
$A(x)$ and $B_c(x)$ have a second root, common to both, but this root is not in the interval $[0,1]$ and therefore not of interest.\\
As noted previously, necessary and sufficient conditions for these expressions to correspond to a unique BNE are $b_+\leq s_+$ and $a_- \geq d_-$. These conditions are equivalent with
\begin{align}
\label{alfbet}
\frac{d_+-s_-}{(\sqrt{\alpha}+\sqrt{\beta})^2}\leq \frac{\sqrt{\alpha}}{ \sqrt{\alpha}+2\sqrt{\beta}} \, , \quad \frac{d_+-s_-}{(\sqrt{\alpha}+\sqrt{\beta})^2}\leq \frac{\sqrt{\beta}}{ \sqrt{\beta}+2\sqrt{\alpha}} \, .
\end{align}
We note that not every pair $(\alpha,\beta) \in [0,1]^2$ yields a supply and demand function that intersect in a point with $x$-coordinate between one and zero. It is easy to check that we need $\alpha+\beta\geq d_+-s_-$. \\
Using the expression for $A(x)$ and (\ref{defA2}), we find in the case $\alpha \neq \beta$ the following expression from which $a(m)$ can easily be recovered:
\begin{align}
&\frac{(\alpha-\beta)}{(1-\gamma)}\frac{(a(m)-s_-)}{(d_+-s_-)}= \nonumber \\
&\alpha+\beta+\sqrt{\alpha \beta}-(\beta+2\sqrt{\alpha \beta})\sqrt{1-\frac{4}{(2\sqrt{\alpha}+\sqrt{\beta})^2}\frac{(\alpha-\beta)}{(1-\gamma)}\frac{(m-s_-)}{(d_+-s_-)}} \, ,
\end{align}
and when $\alpha=\beta$, we have
\begin{align}
\label{asym}
a(m)=\frac{2}{3}m+\frac{1}{4}d_++\frac{1}{12}s_- \, .
\end{align}
In both cases these expressions hold for all $s_-\leq m \leq b_+$. It is straightforward to check that, indeed, $a(s_-)=a_-$ and $a(b_+)=b_+$.\\
Similarly, we have in the case $\alpha \neq \beta$,
\begin{align}
&\frac{(\alpha-\beta)}{\gamma}\frac{(d_+-b(M))}{(d_+-s_-)}= \nonumber \\
&-(\alpha+\beta+\sqrt{\alpha \beta})+(\alpha+2\sqrt{\alpha \beta})\sqrt{1+\frac{4}{(\sqrt{\alpha}+2\sqrt{\beta})^2}\frac{(\alpha-\beta)}{\gamma}\frac{(d_+-M)}{(d_+-s_-)}} \, ,
\end{align}
and when $\alpha=\beta$, we have
\begin{align}
\label{bsym}
b(M)=\frac{2}{3}M+\frac{1}{4}s_-+\frac{1}{12}d_+ \, .
\end{align}
In the special case that the types of both sellers and buyers are uniformly distributed over $[0,1]$, so that $S(x)=x$ and $D(x)=1-x$, we have
\begin{align}
\label{symBNE}
a(m)&=\frac{2}{3}m+\frac{1}{4} \quad , \quad  0\leq m \leq \frac{3}{4} \nonumber \\
b(M)&=\frac{2}{3}M+\frac{1}{12} \quad , \quad  \frac{1}{4}\leq M \leq 1 \, .
\end{align}
There is an interesting connection with the $k$-double auction, introduced by Chatterjee and Samuelson (1983), particularly with $k=1/2$. In this auction, one unit of a good is auctioned. There is a buyer, whose private valuation of the good to be auctioned is a random variable whose probability distribution function is common knowledge. Similarly, there is a seller who also has a private valuation of the unit, again a random variable with a commonly known probability distribution function. First, Nature decides the valuations for the buyer and the seller, after which the buyer proposes a sealed bid and the seller a sealed ask. If the bid is equal or higher than the ask, a transaction occurs with transaction price in the case $k=1/2$ equal to the average of the bid and the ask. A strategy for a buyer is a function that assigns to each valuation a bid and similarly for the seller. The problem is to find pairs of strategy functions that form a Bayesian Nash equilibrium. \\
A slightly different interpretation is that there are infinite sets of buyers and sellers. The distribution of their types is given by the commonly known probability density, which now serves as a distribution function. Nature randomly chooses a buyer and a seller from these populations. Traders of the same type will play the same strategy, i.e. the bid they will offer or the ask they will make. After the buyer and seller are chosen, the corresponding bid and ask are compared and if the bid is higher or equal to the ask, there is a transaction with price equal to the average of the ask and the bid. Again, we are looking for Bayesian Nash equilibria.\\ 
Chatterjee and Samuelson found that in the case that the valuations are uniformly distributed over the interval $[0,1]$, for both the sellers and the buyers, a Bayesian Nash strategy is given by expression (\ref{symBNE}), exactly as in this game. This might suggest that the solutions found for this game might also be solutions of the $k$-double auction, but it is relatively easy to check that this is not the case.\\
Satterthwaite and Williams (1989) showed that, in contrast to this game, Bayesian Nash equilibria of the $k$-double auction are generally not unique, but form a two-parameter continuous family of equilibria.
\section{One-price strategies}
\begin{figure}
\centering
\subfloat[][]{\includegraphics[scale=0.35]{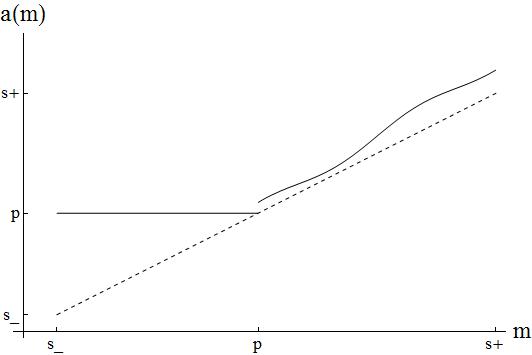}}
\qquad
\subfloat[][]{\includegraphics[scale=0.35]{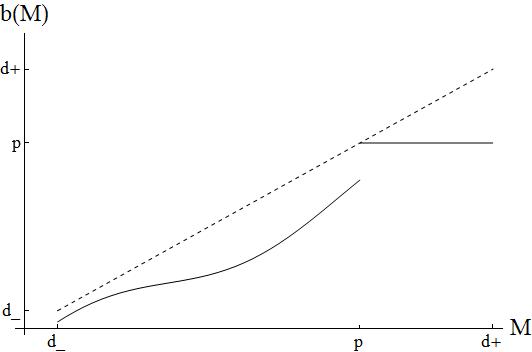}}
\caption{The strategies $a(m)$ and $b(M)$.}
\label{strategies}
\end{figure}

\begin{figure}
\centering
\subfloat[][]{\includegraphics[scale=0.35]{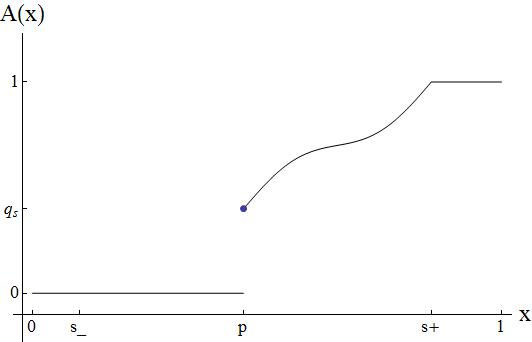}}
\qquad
\subfloat[][]{\includegraphics[scale=0.35]{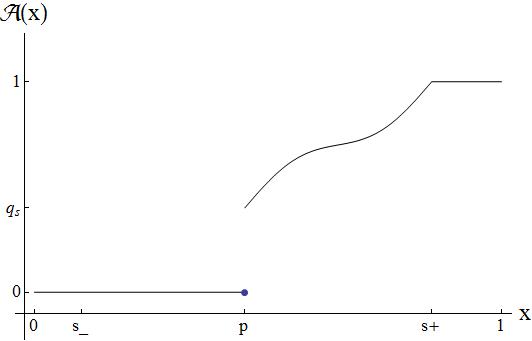}}
\caption{Cumulative distribution of asks $A(x)$ and ${\cal A}(x)$.}
\label{cumdisA}
\end{figure}

\begin{figure}
\centering
\subfloat[][]{\includegraphics[scale=0.35]{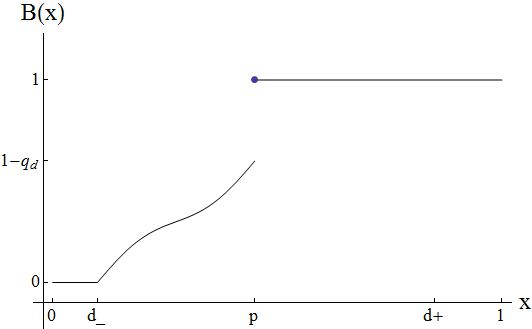}}
\qquad
\subfloat[][]{\includegraphics[scale=0.35]{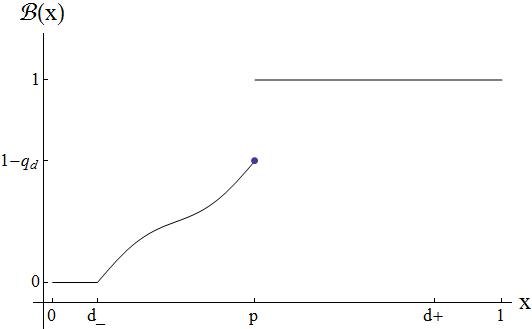}}
\caption{Cumulative distribution of bids $B(x)$ and ${\cal B}(x)$.}
\label{cumdisB}
\end{figure}

We will define a {\it one-price strategy}, with price $p$, as a strategy such that every buyer of type greater or equal to $p$ will bid $p$, while every seller of type less or equal to $p$ will ask $p$. As it is clear that all transactions will happen at a price $p$, those buyers whose type is below this value are never involved, nor are the sellers with a type greater than $p$; they are extramarginal traders.\\
The strategies of the extramarginal traders are inconsequential, but for simplicity we will restrict them to a.e. continuously differentiable ask- and bid functions. A pair of one-price strategies is therefore given by 
\begin{align*}
a(m)=\begin{cases} p, & \mbox{if } m\in [s_-,p] \\ \bar{a}(m), & \mbox{if } m\in (p,s_+] \end{cases} \, ,
\end{align*}
and
\begin{align*}
b(M)=\begin{cases}  \bar{b}(M), & \mbox{if } M\in [d_-,p)  \\ p, & \mbox{if } M\in [p,d_+]\end{cases} \, .
\end{align*}
Here $\bar{a}(m)$ is a continuously differentiable function such that $\bar{a}(m) \geq m$ and $\bar{b}(M)$ is a continuously differentiable function such that $\bar{b}(M) \leq M$, see Figure \ref{strategies}.\\
We will calculate the payoff functions associated with one-price strategies and show that none of them form a BNE.\\
Assume that $p\geq p^*$, with $p^*$ the competitive price. To this price $p$ there correspond fractions $q_d<q_s$ such that $S(q_d)=p$ and $D(q_d)=p$. We have that 
\begin{align*}
&A(x)=0 \, , \, s_-\leq x<p \quad , \quad A(p)=q_s \\
&{\cal A}(x)=0 \, , \, s_-\leq x\leq p\\
&B(x)=1 \, , \, p \leq x< d_+  \\
&{\cal B}(x)=1 \, , \, p < x\leq d_+ \quad ,\quad {\cal B}(p)=1-q_d \, .
\end{align*}
The graphs of the above expressions are shown in Figures \ref{cumdisA} and \ref{cumdisB}.\\

\begin{figure}[tb]
\centering
\includegraphics[scale=0.45]{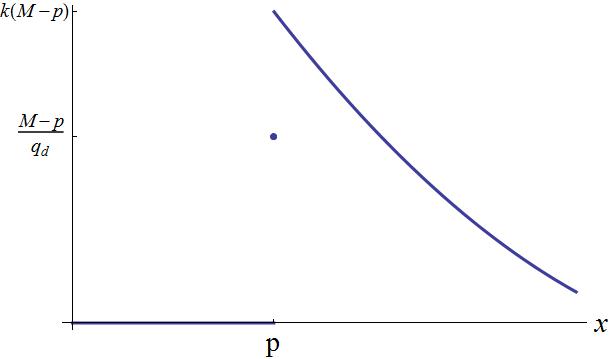}
\caption{Graph of $\pi_b(x,M)$.} 
\label{pi1price}
\end{figure}

Using (\ref{gamma12}), (\ref{gamma22}) and (\ref{pib2}), we find for $x<p$
\begin{align*}
\pi_b(x,M)=0 \,  ,
\end{align*}
where we have used a shorthand for $\pi_b(x,M;a(.),b(.))$. When $x=p$, we have
\begin{align*}
\pi_b(p,M)&=\frac{(M-p)q_s}{(q_s+q_d)q_s}+\frac{(M-p)q_s}{(q_s+q_d)q_d}\\
&=\frac{M-p}{q_d} \, .
\end{align*}
Finally, for $p<x\leq d_+$ we find
\begin{align*}
\pi_b(x,M)&=\frac{M-x}{A(x)}+\frac{(M-p)q_s}{(q_s+q_d)q_d}+\int_p^x(M-q)\frac{A'(q}{A(q)^2}\, dq \\
&=\frac{(M-p)q_s}{(q_s+q_d)q_d}+\frac{M-p}{A(p)}-\int_p^x\frac{1}{A(q)}\, dq \\
&=k(q_s,q_d)(M-p)-\int_p^x\frac{1}{A(q)}\, dq \, ,
\end{align*}
with
\begin{align*}
k(q_s,q_d)=\frac{q_s^2+q_d^2+q_s q_d}{q_s q_d(q_s+q_d)}
\end{align*}
It is clear that $\pi_b(x,M)$ is discontinuous in $x=p$. We have that
\begin{align*}
k(q_s,q_d)-\frac{1}{q_d}&=\frac{q_s^2+q_d^2+q_s q_d}{q_s q_d(q_s+q_d)}-\frac{1}{q_d} \\
&=\frac{q_d}{q_s(q_s+q_d)}>0 \, .
\end{align*}
The graph of $\pi_b(x,M)$ is shown in Figure \ref{pi1price}. We conclude from this figure that when all sellers use strategy $a(m)$ and all other buyers use strategy $b(M)$, then for a buyer with limit $M>p$, it is profitable to play a strategy larger than $p$. Although there does not exist a unique best reply, there is a whole interval of values of $x$ that lead to a higher profit for this buyer than playing $x=p$.\\
This result is not surprising. When all colleagues/competitors on the demand side who can afford it, bid the same price, bidding slightly higher than this price becomes profitable. The profit margin decreases somewhat, but the increase in probability of winning, more than makes up for that. A similar calculation can be made for the supply side, which shows that asking slightly less is profitable for a seller, when all other sellers ask the same price.\\
It is worth noting that the relative profit jump for buyers
\begin{align*}
\frac{1}{M-p}(\lim_{x \downarrow p}\pi_b(x,M)-\pi_b(p,M))=\frac{q_d}{q_s(q_s+q_d)}
\end{align*}
is equal to the relative profit jump for sellers
\begin{align*}
\frac{1}{p-m}(\lim_{x \uparrow p}\pi_a(x,m)-\pi_a(p,m))=\frac{q_s}{q_d(q_s+q_d)} \, ,
\end{align*}
if and only if $q_s=q_d$, i.e. if $p=p^*$.

\section{Comparison of the BNE and competitive behaviour for linear supply and demand}
Many interesting aspects of the BNE can be calculated explicitly in the  case of linear supply and demand functions. Here, we will focus on two of these, namely the fraction of intramarginal traders and the sum of the expected profit of the successful seller and buyer, which we will call total expected profit. These will be compared to the case that the players play competitive strategies.\\
The auction where the players use competitive strategies will be called the {\it competitive market}. In this market, every buyer of type greater or equal to $p^*$ and every seller of type less or equal to $p^*$ are intramarginal, so for both the populations of buyers and sellers, the fraction of intramarginals is $q^*$. For linear supply and demand functions $S(x)=s_-+\alpha x$ and  $D(x)=d_+-\beta x$, these values are
\begin{align*}
p^*=\frac{\alpha d_++\beta s_-}{\alpha+\beta}=s_-+\frac{\alpha (d_+- s_-)}{\alpha+\beta}\quad , \quad q^*=\frac{ d_++ s_-}{\alpha+\beta} \, .
\end{align*}
In the case of the BNE, sellers with a limit less or equal to $b_+$ are intramarginal, as are buyers with a limit greater or equal to $a_-$. The set of intramarginal buyers and sellers is larger for the BNE than for the competitive market. This can be seen by noting that 
\begin{align*}
a_-&=s_-+\frac{\alpha(d_+-s_-)}{(\sqrt{\alpha}+\sqrt{\beta})^2} \\
&=s_-+\frac{\alpha (d_+- s_-)}{\alpha+\beta}-\frac{\alpha (d_+- s_-)}{\alpha+\beta}+\frac{\alpha(d_+-s_-)}{(\sqrt{\alpha}+\sqrt{\beta})^2}\\
&=p^*+\alpha (d_+-s_-)(\frac{1}{(\sqrt{\alpha}+\sqrt{\beta})^2}-\frac{1}{\alpha+\beta})<p^* \, .
\end{align*}
Similarly, $b_+>p^*$.\\
\\
When the sellers and buyers use strategy profiles $a(m)$ and $b(M)$, respectively, the profit density for the sellers is given by $\pi_a(x=a(m),m)$. This means that the fraction of sellers whose limit is $m \in[m_1,m_2]$ have an expected profit of $\int_{m_1}^{m_2}\sigma(m)\pi_a(x=a(m),m)\, dm$. Note that $\pi_a(x=a(m),m)$ also depends on $b(M)$.\\
The expected payoff for the whole population of sellers is given by
\begin{align}
\label{profa}
P_a=\int_{s_-}^{s_+} \sigma(m)\pi_a(x=a(m),m)\, dm \, .
\end{align}
In the case of the competitive market, it follows from the previous section that
\begin{align*}
\pi_a(x=a(m),m)=\begin{cases}  \frac{(p^*-m)}{q^*}, & \mbox{if } m\in [s_-,p^*]\\ 0, & \mbox{if } m\in (p^*,s_+] \end{cases} \, ,
\end{align*}
so that 
\begin{align*}
P_a^c=\int_{s_-}^{p^*}\frac{\sigma(m)(p^*-m)}{q^*}\, dm \, =\frac{1}{q^*}\int_{0}^{q^*}(p^*-S(x))dx\, , 
\end{align*}
where the last equation results from the substitution $x=S^{-1}(m)$. This expression is easily calculated:
\begin{align*}
P_a^c=\frac{1}{2}(p^*-s_-) \, . 
\end{align*}
By a similar reasoning we find that the expected profit for the buyers in the competitive market is given by
\begin{align*}
P_b^c=\frac{1}{2}(d_+-p^*) \, , 
\end{align*}
so that the total expected profit in the competitive market is
\begin{align}
\label{profcomp}
P^c=\frac{1}{2}(d_+-s_-) \, . 
\end{align}
For the case of the BNE, (\ref{profa}) becomes:
\begin{align}
\label{profaBNE}
P_a^{BNE}=\int_{s_-}^{b_+}\!\!\!\sigma(m)\left(\frac{(a(m)-m)B_c(a(m))}{(A(a(m))+B_c(a(m)))^2}-\int_{a(m)}^1\frac{(q-m)}{(A(q)+B_c(q))^2}B_c'(q) dq\right) dm
\end{align}
Because the functions in the above expression are known and relatively simple, this integral can be calculated. In Appendix \ref{totprof}, it is shown that, quite surprising, also here
\begin{align*}
P^{BNE}=\frac{1}{2}(d_+-s_-)\, ,
\end{align*}
or in other words:
\begin{align*}
P^{BNE}=P^c\, .
\end{align*}
\section{Conclusion}
We have introduced a game which models a simple version of a continuous
double auction. The players of the game are either buyers or sellers of one
good and each buyer has his own maximum buying price and each seller a minimum selling
price. The aggregation of these maxima and minima produce demand
 and supply functions, respectively, the intersection of which define
the competitive equilibrium. The players choose a buying price (buyers) or
selling price (sellers) before the start of the game. Then, a unit of the good
is auctioned using the continuous time double auction method.\\
Although this model is a very reduced form of real continuous double auctions,
it retains the important feature that in choosing strategies, players have
to balance the desire for a high profit with the need to outbid or undersell
the competition.\\
The functions describing the expected payoffs were found to have a quite simple
form. Also, the distribution of the expected transaction prices is given by
a simple formula, which as a by-product also serves as an expression for the
price of the first transaction in Gode and Sunder's ZI-C model. A further
analysis of the payoffs of this game showed that for linear supply and demand
functions, there exists either no Bayesian Nash equilibrium within the set of
differentiable and strictly increasing strategies, or a unique one, and in the
latter case an explicit expression was found.\\
It was shown that the set of strategies corresponding to a market in competitive
equilibrium is not a Bayesian Nash equilibrium.\\
Finally, it was found that the Bayesian Nash solution has a welfare property
in common with the competitive market, namely the sum of the expected
payoffs for sellers and buyers is the same in both markets. This is illustrated
in Figure 10, which shows the profit densities for the sellers in the case of
the Bayesian Nash equilibrium and the case of the competitive market, for
the uniform case where the demand function is $D(x) = 1-x$ and the supply
function is $S(x) = x$, both on $[0, 1]$. Because the supply and the demand
functions are symmetric, the profit densities for the buyers are similar. The
expected profit for the sellers is equal to the integral of the function, which
in both cases equals $0.25$. For the competitive market this is obvious. The
first seller chosen with a limit price equal or lower than the competitive price
$p^* = 0.5$ is the one involved in the transaction and she will make a profit
equal to $p^*$ minus her limit price. Since the limit prices of the sellers are
uniformly distributed over $[0,1]$, it is clear that the average profit will be
$0.25$. The expected profit in the case of the Bayesian Nash equilibrium can
be calculated, and due to the symmetry of the supply and demand function
the calculation only involves integrating a quadratic function.

\begin{figure}[tb]
\centering
\includegraphics[scale=0.5]{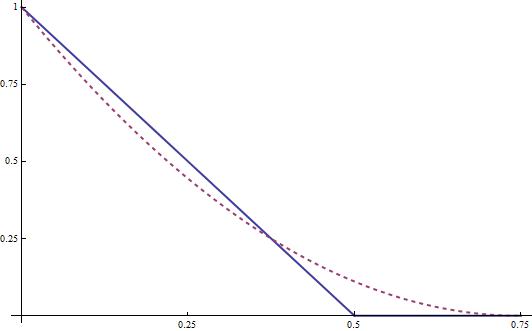}
\caption{Profit density function $\pi_a(x = a(m);m)$ for the Bayesian Nash
equilibrium (dashed line) and for the competitive strategy (solid line), for $S(x)=x$ and $D(x)=1-x$.} 
\label{}
\end{figure}

There are still many unanswered questions about this game. Are there
Bayesian Nash equilibria which are not differentiable and strictly increasing?
Does the uniqueness of the Bayesian Nash equilibrium and its welfare
property also hold if the supply and demand function are not linear? Can
the results be translated to a setting with a finite number of players?\\
Probably the most important unsolved problem is to give a strategic foundation
for competitive equilibrium in the type of double auctions that have
been so widely studied since Smith (1962). The next step in this line of research will have
to be to expand the game to include multiple rounds, with successful traders
exiting the market after each round.

\appendix
\section{Derivation of the distribution of outcomes}\label{derivout}
To translate (\ref{betaexpr}) into a mathematical expression, we will introduce some notation. Let $X_k$ denote the stochastic variable defined by the maximum bid at $\tau=k$ and $Y_k$ the minimum ask at that time. Let $typemax_k$ be the type of the buyer who holds the maximum bid at $\tau=k$, and $typemin_k$ the type of the seller who holds the minimum ask at that time.\\
Using these definitions, and recalling that a buyer of type $M$ uses strategy $x$, we can write:
\begin{align}
\label{notrans}
&Pr({\rm no\: transaction\: at\:} \tau=n-1 {\rm \:and\: holder\: maximum\: bid \:is \:of \: type\:} M)\,{\rm d}M = \nonumber \\
&Pr(X_{n-1}=x \: {\rm and\:} Y_{n-1}>x \:{\rm and\:} typemax_{n-1}=M) {\rm d}M\, .
\end{align}
The event $(X_{n-1}=x,Y_{n-1}>x , typemax_{n-1}=M)$ can occur in a number of ways, which we classify according to the number of bids in the sequence of $n-1$ shouts made up till time $\tau=n-1$.\\
Since $X_{n-1}=x>0$, it follows that at $\tau=n-1$ at least one bid was made. Let $k\geq 1$ be the number of bids in the sequence, then there are $n-1-k$ asks in the sequence, each of which is larger than $x$. The probability that $n-1-k$ sellers were chosen and that all asks are larger than $x$ is $((1-A(x))/2)^{n-1-k}$. There are $\binom{n-1}{n-1-k}$$=\binom{n-1}{k}$ ways in which the asks can be distributed over the $n-1$ positions of the sequence.\\
Of the $k$ bids, one must be made by a buyer of type $M$, which happens with probability(density) $\mu(M){\rm d}M$. The other $k-1$ bids must either be all smaller than $x$, or the bids that are equal to $x$ must have been made after the buyer of type $M$ had made his bid. If we consider the sub-sequence of bids, the winning bid can be in position $1$, $2$,$\ldots$, $k$. Therefore, the probability(density) that $k$ buyers were selected and that in the subsequence of bids a player of type $M$ was chosen, all buyers before him made a bid smaller than $x$ and all those after him a bid smaller or equal to $x$ is:
\begin{align}
(\frac{1}{2})^k\sum_{l=1}^k {\cal B}(x)^{l-1}B(x)^{k-l}\mu(M){\rm d}M\, .
\end{align}
Putting the above together, we find:
\begin{align}
\label{notrans2}
&Pr(X_{n-1}=x \: {\rm and\:} Y_{n-1}>x \:{\rm and\:} typemax_{n-1}=M){\rm d}M\,  = \nonumber \\
&(\frac{1}{2})^{n-1}\mu(M)\sum_{k=1}^{n-1}\binom{n-1}{k}(1-A(x))^{n-1-k}\sum_{l=1}^k {\cal B}(x)^{l-1}B(x)^{k-l} \, {\rm d}M \nonumber \\
&=u_{n-1}(x)\mu(M){\rm d}M
\end{align}
Expression (\ref{notrans2}) can be simplified. There are two cases:
\begin{enumerate}
\item ${\cal B}(x)=B(x)$. Then $\displaystyle\sum_{l=1}^k {\cal B}(x)^{l-1}B(x)^{k-l}=k\, B(x)^{k-1}$.\\
Since for general $v, w \in {\mathbb R}$, $\displaystyle\sum_{k=1}^N \binom{N}{k}k\,v^{N-k}w^{k-1}=N(v+w)^{N-1}$, we find that
\begin{align}
u_{n-1}(x)=\frac{1}{2}(n-1)((1-A(x)+B(x))/2)^{n-2}\, .
\end{align}
\item ${\cal B}(x)<B(x)$. Then $\displaystyle\sum_{l=1}^k {\cal B}(x)^{l-1}B(x)^{k-l}$$=B(x)^{k-1}\displaystyle\sum_{l=1}^k (\frac{{\cal B}(x)}{B(x)})^{l-1}$\\
=$\dfrac{B(x)^{k}}{B(x)-{\cal B}(x)}(1-(\dfrac{{\cal B}(x)}{B(x)})^{k})$. Substitution yields
\begin{align}
u_{n-1}(x)=&\frac{1}{B(x)-{\cal B}(x)}\big( \sum_{k=1}^{n-1}\binom{n-1}{k}((1-A(x))/2)^{n-1-k}(B(x)/2)^k- \nonumber\\
&\sum_{k=1}^{n-1}\binom{n-1}{k}((1-A(x))/2)^{n-1-k}({\cal B}(x)/2)^k \big) \nonumber\\
=&\frac{1}{B(x)-{\cal B}(x)}\big( ((1-A(x)+B(x))/2)^{n-1}-\nonumber \\
&((1-A(x)+{\cal B}(x))/2)^{n-1}\big).
\end{align}
\end{enumerate}
The other factors in (\ref{betaexpr}) have a straightforward expression:
\begin{align}
\label{otherfacs}
&Pr({\rm seller\: with\: limit\:} m {\rm \:is \:chosen})\,{\rm d}m=\frac{1}{2}\sigma(m)\,{\rm d}m \nonumber \\
&Pr(a(m)\leq x)=\theta(x-a(m)) \nonumber \\
&Pr(t=x)\,{\rm d}t=\delta (t-x)\,{\rm d}t\, ,
\end{align}
with $\delta(z)$ the Dirac delta distribution.\\
\\
Substituting (\ref{notrans2}) and (\ref{otherfacs}) in the first part of (\ref{betaexpr}), we find:
\begin{align*}
&\sum\limits_{n=2}^\infty Pr({\rm no\: transaction\: at\:} \tau=n-1 {\rm \:and\: holder\: maximum\: bid\:is \:of \: type\:} M)\\
&Pr({\rm seller\: with\: limit\:} m {\rm \:is \:chosen})\, Pr(a(m)\leq x)\, Pr(t=x){\rm d}M\, {\rm d}m\, {\rm d}t\, = \\
&\frac{1}{2}\left(\sum\limits_{n=2}^\infty u_{n-1}(x)\right)\mu(M)\sigma(m)\theta(x-a(m))\delta(t-x) {\rm d}M\, {\rm d}m\, {\rm d}t \, =\\
&\gamma_1(x)\mu(M)\sigma(m)\theta(x-a(m))\delta(t-x) {\rm d}M\, {\rm d}m\, {\rm d}t \,
\end{align*} 
Again we consider two cases:
\begin{enumerate}
\item ${\cal B}(x)=B(x)$. Then 
\begin{align}
\label{beta11}
&\gamma_1(x)=\frac{1}{4}\sum\limits_{n=2}^\infty (n-1)((1-A(x)+B(x))/2)^{n-2}\, = \nonumber \\
&\frac{1}{4}\frac{1}{(1-((1-A(x)+B(x))/2))^2}=\frac{1}{(1-B(x)+A(x))^2}\, \, .
\end{align}
\item ${\cal B}(x)<B(x)$. Then 
\begin{align}
\gamma_1(x)=&\frac{1}{2}\sum\limits_{n=2}^\infty\frac{1}{B(x)-{\cal B}(x)}\big( ((1-A(x)+B(x))/2)^{n-1}-\nonumber \\
&((1-A(x)+{\cal B}(x))/2)^{n-1}\big) \nonumber\\
=&\frac{1}{2(B(x)-{\cal B}(x))}\big(\frac{1-A(x)+B(x)}{1-B(x)+A(x)}-\frac{1-A(x)+{\cal B}(x)}{1-{\cal B}(x)+A(x)}\big) \nonumber \\
=&\frac{1}{(1-B(x)+A(x))(1-{\cal B}(x)+A(x))}.
\end{align}
\end{enumerate}
In an analogous fashion we find for the second part of (\ref{betaexpr}):
\begin{align}
&Pr({\rm no\: transaction\: at\:} \tau=n-1 {\rm \:and\: holder\: minimum\: ask\: is \:of\: type\:} m)= \nonumber \\
&Pr(X_{n-1}<a(m) \: {\rm and\:} Y_{n-1}=a(m) \:{\rm and\:} typemin_{n-1}=m) {\rm d}m \, .
\end{align}
Therefore,
\begin{align}
&\sum\limits_{n=2}^\infty Pr({\rm no\: transaction\: at\:} \tau=n-1 {\rm \:and\: holder\: minimum\: ask\: is \:of\: type\:} m)\nonumber\\
&Pr({\rm buyer\: with\: limit\:} M {\rm \:is \:chosen})\, Pr(a(m)\leq x)\, Pr(t=a(m))\, = \nonumber\\
&\sum\limits_{n=2}^\infty Pr(X_{n-1}<a(m) \: {\rm and\:} Y_{n-1}=a(m) \:{\rm and\:} typemin_{n-1}=m)\, = \nonumber\\
&\gamma_2(a(m))\,\mu(M)\sigma(m)\theta(x-a(m))\delta(t-a(m)) {\rm d}M\, {\rm d}m\, {\rm d}t \,
\end{align}
with
\begin{enumerate}
\item ${\cal A}(y)=A(y)$. Then 
\begin{align}
&\gamma_2(y)=\frac{1}{(1-{\cal B}(y)+A(y))^2}\, \, .
\end{align}
\item ${\cal A}(y)<A(y)$. Then 
\begin{align}
&\gamma_2(y)=\frac{1}{2(A(y)-{\cal A}(y))}\big(\frac{1-{\cal A}(y)+{\cal B}(y)}{1-{\cal B}(y)+{\cal A}(y)}-\frac{1-A(y)+{\cal B}(y)}{1-{\cal B}(y)+A(y)}\big) \nonumber\\
=&\frac{1}{(1-{\cal B}(y)+A(y))(1-{\cal B}(y)+{\cal A}(y))}.
\end{align}
\end{enumerate}
Putting it al together we find:
\begin{align}
&\beta(M,m,t;x,a(m),b(M))\,= \nonumber \\
&\mu(M)\sigma(m)\theta(x-a(m))\left(\gamma_1(x)\delta(t-x)+\gamma_2(a(m))\delta(t-a(m))\right)\, .
\end{align}
By an analogous reasoning, we find
\begin{align}
&\alpha(M,m,t;x,a(m),b(M))\,= \nonumber \\
&\mu(M)\sigma(m)\theta(b(M)-x)\left(\gamma_2(x)\delta(t-x)+\gamma_1(b(M))\delta(t-b(M))\right)\, .
\end{align}

\section{Proof of Lemma 2}\label{proof}
Let $P_{\sigma}$ be the measure on $[0,1]$ which for every measurable set $B\subset [0,1]$ is defined by:
\begin{align*}
P_{\sigma}(B)=\int_B \sigma(m)\, {\rm d}m
\end{align*}
Then
\begin{align*}
A(q)&:=\int_{s_-}^{s_+} \theta(q-a(m))\sigma(m)\, {\rm d}m=P_{\sigma}(\{m\in [s_-,s_+]\, | \, q \geq a(m)\}) \\
&=P_{\sigma}(a^{-1}([a_-,q])) \, .
\end{align*}
We find that 
\begin{align}
\int_{s_-}^{s_+} g(t)\sigma(m)f(a(m))\delta(t-a(m))\, {\rm d}t\, {\rm d}m&=\int_{s_-}^{s_+} g(a(m))f(a(m))\sigma(m)\, {\rm d}m \nonumber \\
&=\int_{s_-}^{s_+} g(a(m))f(a(m)){\rm d}P_{\sigma}(m) \label{pullback}
\end{align}
The Transformation lemma for measures (see for instance Halmos [1950], Theorem C, p.163) states that if $\nu({\rm d}x)$ is a measure on $[0,1]$ and $T:{\mathbb R} \rightarrow {\mathbb R}$ is a measurable transformation, then
\begin{align*}
\int_B f(T(x))\nu({\rm d}x)=\int_{T(B)} f(y)\nu_T ({\rm d}y) \, ,
\end{align*}
for every measurable $B\subset [0,1]$ and every measurable function $f:T([0,1]) \rightarrow {\mathbb R}$ and with $\nu^T(C)=\nu(T^{-1}(C))$ for all measurable sets $C\subset T([0,1])$.\\
Applying the Transformation lemma with $T(x)=a(x)$ to (\ref{pullback}), we find
\begin{align*}
\int_{s_-}^{s_+} g(a(m))f(a(m)){\rm d}P_{\sigma}(m) = \int_{a_-}^{a_+} g(q)f(q){\rm d}P_{\sigma}^a(q)
\end{align*}
Since 
\begin{align*}
P_{\sigma}^a([0,q])=P_{\sigma}(a^{-1}([0,q]))=A(q) \, ,
\end{align*}
we have that 
\begin{align*}
\int_{a_-}^{a_+} g(q)f(q){\rm d}P_{\sigma}^a(q)&=\int_{a_-}^{a_+} g(q)f(q){\rm d}A(q) \\
&=\int_0^1 g(q)f(q){\rm d}P_{\sigma}^a(q) \, ,
\end{align*}
where the last equality follows from the fact that $A(x)=0$ for $x\leq a_-$ and $A(x)=1$ for $x\geq a_+$.\\
The proof of (\ref{lemeq2}) is similar, using the fact that
\begin{align}
\label{complbeta}
\int_{d_-}^{d_+}\!\!\theta(b(M)-x)\,\mu(M)\, {\rm d}M=1-{\cal B}(x) \, .
\end{align}
\section{Derivation of the expression for the transaction prices}\label{trans}
\begin{align*}
T(t)&=\int_0^t\int_{d_-}^{d_+} \int_{s_-}^{s_+}\!\!\! \beta(M,m,t';x=b(M),a(m),b(M))\, {\rm d}m{\rm d}M {\rm d}t' \\
&=\int_0^t\int_{d_-}^{d_+} \int_{s_-}^{s_+}\!\!\!\mu(M)\sigma(m)\theta(b(M)-a(m))\, \big (\gamma(b(M))\delta(t'-b(M))\\
&+\gamma(a(m))\delta(t'-a(m))\big )\, {\rm d}m{\rm d}M {\rm d}t'\\
&=\int_0^t\int_{d_-}^{d_+}\!\!\!\mu(M)\gamma(b(M))\delta(t'-b(M))\!\!\int_{s_-}^{s_+}\!\!\!\sigma(m)\theta(b(M)-a(m)){\rm d}m\,{\rm d}M {\rm d}t'\\
&+\int_0^t\int_{s_-}^{s_+}\!\!\!\sigma(m)\gamma(a(m))\delta(t'-a(m))\!\!\int_{d_-}^{d_+}\!\!\!\mu(M)\theta(b(M)-a(m)){\rm d}M\,{\rm d}m {\rm d}t'\\
&=\int_0^t\int_{d_-}^{d_+}\!\!\!\mu(M)\gamma(b(M))\delta(t'-b(M))A(b(M)){\rm d}M {\rm d}t'\\
&+\int_0^t\int_{s_-}^{s_+}\!\!\!\sigma(m)\gamma(a(m))\delta(t'-a(m))(1-B(a(m)))\,{\rm d}m {\rm d}t' \\
&=\int_0^1\int_{d_-}^{d_+}\!\!\!\theta(t-t')\mu(M)\gamma(b(M))\delta(t'-b(M))A(b(M)){\rm d}M {\rm d}t'\\
&+\int_0^1\int_{s_-}^{s_+}\!\!\!\theta(t-t')\sigma(m)\gamma(a(m))\delta(t'-a(m))(1-B(a(m)))\,{\rm d}m {\rm d}t'
\end{align*}
Applying equation (\ref{lemeq2}) to the first part of the equality and (\ref{lemeq1}) to the second part then yields
\begin{align}
T(t)&=\int_0^t \gamma(t')(A(t')B'(t')+(1-B(t'))A'(t'))\,{\rm d}t'\nonumber \\
&=\int_0^t \frac{(A(t')B'(t')+(1-B(t'))A'(t'))}{(1-B(t')+A(t'))^2}\,{\rm d}t'\nonumber \\
&=\int_0^t \frac{{\rm d}}{{\rm d}t'}\frac{A(t')}{(1-B(t')+A(t'))}\,{\rm d}t'\nonumber \\
&=\frac{A(t)}{(1-B(t)+A(t))}\, .
\end{align}
\section{Solution for linear supply and demand functions}\label{solution}
Recalling the definition
\begin{align*}
T=\frac{A}{A+B_c} \, ,
\end{align*}
the equation for $T$ becomes:
\begin{align}
\label{Teq}
2(d_+-\beta B_c-x)T'-T=0 \, ,
\end{align}
with $T(a_-)=0$ and $T(b_+)=1$. From (\ref{condB}) we find:
\begin{align}
\label{Beq}
(\alpha T^2-\beta(1-T)^2)B_c=(1-T)(x-(T s_-+(1-T)d_+)) \, .
\end{align}
Note that this equation is satisfied for all $B_c$ and $T=\gamma$ and $x=\bar{x}$, with
\begin{align*}
\gamma=\frac{\sqrt{\beta}}{\sqrt{\alpha}+\sqrt{\beta}}\, , \quad \bar{x}=\gamma s_-+(1-\gamma)d_+ \, .
\end{align*}
The above expression does not constitute a solution of (\ref{Teq}), or (\ref{eqlin}) and (\ref{condB}), as it is only defined in $x=\bar{x}$. However, it plays an important role as one of the separatrices of the flow on the surface $V$ defined in the previous section.\\
Rearranging (\ref{Beq}), we find:
\begin{align}
\label{Beq2}
(T-\gamma)((\alpha-\beta)T+\gamma \alpha +(2-\gamma)\beta)B_c=(1-T)(x-\bar{x}+(T-\gamma)(d_+-s_-)) \, ,
\end{align}
which yields
\begin{align}
\label{faceq}
d_+-\beta B_c-x=\frac{(-x(\alpha-\beta)-\beta s_-+\alpha d_+)T^2-\beta(x-s_-)T}{\alpha T^2-\beta(1-T)^2} \, .
\end{align}
Substituting (\ref{faceq}) in (\ref{Teq}), dividing by $T$ and multiplying by ${\alpha T^2-\beta(1-T)^2}$ yields a differential equation which we write in the form
\begin{align}
\label{diffeq}
f(x,T)dT+g(x,T)dx=0
\end{align}
with
\begin{align*}
&f(x,T)=2(\alpha(d_+-x)T+\beta (s_--x)(1-T))\\
&g(x,T)=-\alpha T^2+\beta(1-T)^2 \, .
\end{align*}
We observe that (\ref{diffeq}) is exact, i.e., 
\begin{align*}
\frac{\partial f}{\partial x}=\frac{ \partial g}{\partial T} \, ,
\end{align*}
so that (\ref{diffeq}) can be written in the form 
\begin{align*}
d V(x,T)=0 \, ,
\end{align*}
with
\begin{align*}
V(x,T)=\alpha (d_+-x)T^2-\beta(s_--x)(1-T)^2 \, .
\end{align*}
It follows that we can write
\begin{align}
\label{poteq}
V(x,T)=c \, .
\end{align}
We will take $c$ equal to the value $V(x,T)$ has in the point $(\bar{x},\gamma)$:
\begin{align*}
c=V(\bar{x},\gamma)=\frac{\beta \alpha}{(\sqrt{\alpha}+\sqrt{\beta})^2}(d_+-s_-) \, .
\end{align*}
Equation (\ref{poteq}) can be factorized:
\begin{align*}
&V(x,T)-c=\\
&(T-\gamma)((\alpha(d_+-x)+\beta(x-s_-))T+\gamma \alpha (d_+-x)-(2-\gamma)\beta (x-s_-)=0 \, ,
\end{align*}
giving two branches which intersect in $(\bar{x},\gamma)$ which is therefore a critical point of the flow. These two branches are the separatrices. As noted, $T=\gamma$ does not yield a solution that interests us, so that the the equation for $T(x)$ becomes:
\begin{align}
\label{Texpr}
T(x)=\frac{-\gamma \alpha (d_+-x)+(2-\gamma)\beta (x-s_-)}{\alpha(d_+-x)+\beta(x-s_-)} \, .
\end{align}
After some manipulation, we find
\begin{align}
\label{xbar} 
x-\bar{x}=\frac{\alpha(d_+-x)+\beta(x-s_-)}{2(\gamma \alpha+(1-\gamma)\beta}(T(x)-\gamma) \, .
\end{align}
Substituting (\ref{Texpr}) and (\ref{xbar}) in (\ref{Beq2}), dividing both sides by $T-\gamma$ and some more rearranging, gives the expression:
\begin{align}
&B_c(x)= \nonumber\\
&\frac{(1+\gamma)\alpha (d_+-x)-(1-\gamma)\beta (x-s_-)}{2\beta \alpha (d_+-s_-)}(d_+-s_-+\frac{\alpha(d_+-x)+\beta(x-s_-)}{2\sqrt{\alpha \beta}}). 
\end{align}
Using
\begin{align*}
A(x)=\frac{T(x)}{1-T(x)}B_c(x)\, ,
\end{align*}
we have
\begin{align}
&A(x)= \nonumber\\
&\frac{-\gamma \alpha (d_+-x)+(2-\gamma)\beta (x-s_-)}{2\beta \alpha (d_+-s_-)}(d_+-s_-+\frac{\alpha(d_+-x)+\beta(x-s_-)}{2\sqrt{\alpha \beta}}). 
\end{align}
The expressions for $A(x)$ and $B_c(x)$ are quadratic, but for $\alpha=\beta$, they reduce to linear functions.

\section{Total profit for the BNE strategies}\label{totprof}
In (\ref{profaBNE}) we substitute $x=a(m)$ and note that from (\ref{defA2}) we have $S^{-1}(m)=A(x)$ for all $x \in [s_-,s_+]$, so that $\sigma(m)\, dm=A'(x)\, dx$. Also, $a(s_-)=a_-$ and $a(b_+)=b_+$, and $B'_c(q)=0$, for $q\geq b_+$. The expression (\ref{profaBNE}) then becomes
\begin{align}
\label{profaBNEsub}
P_a^{BNE}=&\int_{a_-}^{b_+}\!\!\!A'(x)\frac{(x-S(A(x)))B_c(x)}{(A(x)+B_c(x))^2}dx-\nonumber \\ &\int_{a_-}^{b_+}A'(x)\int_x^{b_+}\frac{(q-S(A(x)))}{(A(q)+B_c(q))^2}B_c'(q) dq\,dx \, .
\end{align}
Partial integration of the second term of (\ref{profaBNEsub}) yields
\begin{align*}
&\int_{a_-}^{b_+}A'(x)\int_x^{b_+}\frac{(q-S(A(x)))}{(A(q)+B_c(q))^2}B_c'(q) dq\,dx=\\
&\int_{a_-}^{b_+}\!\!\!A(x)\frac{(x-S(A(x)))B'_c(x)}{(A(x)+B_c(x))^2}dx +\\
&\int_{a_-}^{b_+}A(x)(\frac{d\,S(A(x))}{dx})\int_x^{b_+}\frac{B_c'(q)}{(A(q)+B_c(q))^2} dq\, dx \, .
\end{align*}
Substitution in (\ref{profaBNEsub}) produces
\begin{align}
\label{interim}
P_a^{BNE}=&\int_{a_-}^{b_+}\frac{(x-S(A(x)))(A'(x)B_c(x)-A(x)B'_c(x))}{(A(x)+B_c(x))^2}dx\, - \nonumber\\
&\int_{a_-}^{b_+}A(x)(\frac{d\,S(A(x))}{dx})\int_x^{b_+}\frac{B_c'(q)}{(A(q)+B_c(q))^2} dq\, dx \, .
\end{align}
For the BNE, it follows from (\ref{difeqn}) that 
\begin{align}
\label{BNEprop}
2\frac{(x-S(A(x)))(A'(x)B_c(x)-A(x)B'_c(x))}{(A(x)+B_c(x))^2}=\frac{B_c(x)}{A(x)+B_c(x)}=1-T(x)\, .
\end{align}
Substituting (\ref{BNEprop}) in (\ref{interim}), and interchanging the order of integration in the second term finally yields
\begin{align}
\label{finalBNEs}
P_a^{BNE}=\frac{1}{2}\int_{a_-}^{b_+}(1-T(x))\,dx\, - \int_{a_-}^{b_+}\frac{B_c'(q)}{(A(q)+B_c(q))^2}\int_{a_-}^q A(x)(\frac{d\,S(A(x))}{dx})dx\, dq\, .
\end{align}
A similar derivation shows that the expected profit for the buyers is
\begin{align}
\label{finalBNEb}
P_b^{BNE}=\frac{1}{2}\int_{a_-}^{b_+}T(x)\,dx\, + \int_{a_-}^{b_+}\frac{A'(q)}{(A(q)+B_c(q))^2}\int_q^{b_+} B_c(x)(\frac{d\,D(B_c(x))}{dx})dx\, dq\, .
\end{align}
For the total profit we find
\begin{align}
\label{totalBNE}
P^{BNE}&=P_a^{BNE}+P_b^{BNE}=\frac{1}{2}(b_+-a_-) + \nonumber \\
&\int_{a_-}^{b_+}(A(q)+B_c(q))^{-2}(A'(q)\int_q^{b_+} B_c(x)(\frac{d\,D(B_c(x))}{dx})dx\,- \nonumber \\
 &B_c'(q) \int_{a_-}^q A(x)(\frac{d\,S(A(x))}{dx})dx ) dq\, .
\end{align}
In the case of linear supply and demand functions, we have $\frac{d\,S(A(x))}{dx}=\alpha A'(x)$ and $\frac{d\,D(B_c(x))}{dx}=-\beta B'_c(x)$. Substitution in (\ref{totalBNE}) gives
\begin{align}
\label{totalBNE2}
P^{BNE}&=\frac{1}{2}(b_+-a_-) + \frac{1}{2}\int_{a_-}^{b_+}\frac{\beta A'(q)B_c^2(q)- \alpha B_c'(q) A^2(q) }{(A(q)+B_c(q))^{2}} dq \nonumber \\
&=\frac{1}{2}(b_+-a_-) + \frac{1}{2}\int_{a_-}^{b_+}\beta A'(q)(1-T(q))^2- \alpha B_c'(q) T^2(q) dq \, .
\end{align}
We note that 
\begin{align*}
&\int_{a_-}^{b_+} B_c'(q) T^2(q)\, dq =B_c(q) T^2(q)|_{q=a_-}^{q=b_+}-2\int_{a_-}^{b_+} B_c(q) T(q) T'(q)\, dq \\
&=-2\int_{a_-}^{b_+} B_c(q) T(q) T'(q)\, dq=-2\int_{a_-}^{b_+} \frac{A(q)B_c(q)}{A(q)+B_c(q)} T'(q)\, dq \\
&=2\int_{a_-}^{b_+} A(q)(1-T(q))(1- T(q))'\, dq =-\int_{a_-}^{b_+} A'(q)(1-T(q))^2\, dq \, ,
\end{align*}
so that (\ref{totalBNE2}) can be simplified to
\begin{align}
\label{totalBNE3}
P^{BNE}=\frac{1}{2}(b_+-a_-) + (\alpha+\beta)\frac{1}{2}\int_{a_-}^{b_+} A'(q)(1-T(q))^2 dq \, .
\end{align}
We write $P^{BNE}(d_+, s_-, \alpha, \beta)$ to emphasize the dependence of the total profits on the parameters, and observe some facts about this dependency. First, 
\begin{align*}
P^{BNE}(d_+,s_-,k\alpha,k\beta)=P^{BNE}(d_+,s_-,\alpha,\beta) \, ,
\end{align*}
for all $k>0$, so that $P^{BNE}(d_+,s_-,\alpha,\beta)$ only depends on the ratio of $\alpha $ and $\beta$, which we denote by $\frac{\beta}{\alpha}=\lambda^2$. In particular,
\begin{align*}
P^{BNE}(d_+,s_-,\alpha,\beta)=P^{BNE}(d_+,s_-,1,\lambda^2) \, .
\end{align*}
Second, it follows from (\ref{Aexpr}) and (\ref{Texpr}) that 
\begin{align*}
A((d_+-s_-)x+s_-,d_+,s_-,\lambda)&=(d_+-s_-)A(x,1,0,\lambda) \, , \\
T((d_+-s_-)x+s_-,d_+,s_-,\lambda)&=T(x,1,0,\lambda) \, ,
\end{align*}
so that
\begin{align*}
A'((d_+-s_-)x+s_-,d_+,s_-,\lambda)=A'(x,1,0,\lambda) \, .
\end{align*}
Substitution of $q=(d_+-s_-)x+s_-$ in the integral of (\ref{totalBNE3}) produces
\begin{align*}
\int_{a_-}^{b_+} A'(q)(1-T(q))^2 dq =(d_+-s_-)\int_{\frac{a_--s_-}{d_+-s_-}}^{\frac{b_+-s_-}{d_+-s_-}} A'(x,1,0,\lambda)(1-T(x,1,0,\lambda))^2 dx \, .
\end{align*}
Noting that 
\begin{align*}
&b_+-a_-=2\gamma (1-\gamma)(d_+-s_-)=\frac{2\lambda}{(1+\lambda)^2}(d_+-s_-) \, , \\
&\frac{a_--s_-}{d_+-s_-}=\frac{1}{(1+\lambda)^2}=l(\lambda) \, , \quad \frac{b_+-s_-}{d_+-s_-}=\frac{2\lambda+1}{(1+\lambda)^2}=u(\lambda) \, ,
\end{align*}
we see that (\ref{totalBNE3}) can be written as
\begin{align}
\label{totalBNE4}
P^{BNE}=\frac{1}{2}\left(\frac{2\lambda}{(1+\lambda)^2} + (1+\lambda^2)\int_{l(\lambda)}^{u(\lambda)}\! A'(x,\lambda)(1-T(x,\lambda))^2 dx\right)(d_+-s_-)\, .
\end{align} 
Here, the functions in the integrand are obtained from (\ref{Aexpr}) and (\ref{Texpr}), by putting $d_+=1$, $s_-=0$, $\alpha=1$ and $\beta=\lambda^2$. We will assume from here on that $\lambda \neq 1$, as this case will be treated separately.\\
The integrand is a third degree polynomial in $x$, divided by the square of a linear term in $x$. In fact, some computer-assisted manipulation shows that
\begin{align*}
A'(y,\lambda)=\frac{1}{2\lambda^2}((1+\lambda)y+\lambda^2)\, , \quad 1-T(y,\lambda)=\frac{1}{(1-\lambda)y}(y-\frac{2\lambda^2}{1+\lambda}) \, ,
\end{align*}
with
\begin{align*}
y=(\lambda^2-1)x+1 \, .
\end{align*}
Substituting the above expressions in (\ref{totalBNE4}) yields
\begin{align}
\label{totalBNE5}
P^{BNE}=&\frac{1}{2}(\frac{2\lambda}{(1+\lambda)^2} + \nonumber \\
&\frac{(1+\lambda^2)}{2\lambda^2(1-\lambda)^2(\lambda^2-1)}\int_{\hat{l}}^{\hat{u}}(1+\lambda)y-3\lambda^2+\frac{4 \lambda^6}{(1+\lambda)^2}\frac{1}{y^2}\, dy)(d_+-s_-)\, ,
\end{align}
with
\begin{align*}
\hat{l}=\frac{2\lambda}{1+\lambda} \, , \quad \hat{u}=\frac{2\lambda^2}{1+\lambda} \, .
\end{align*}
Note the absence of a $1/y$ term in the integrand in (\ref{totalBNE5}).\\
Evaluating the integral in (\ref{totalBNE5}), also with some help from Mathematica, yields
\begin{align}
\label{totalfac}
&\frac{2\lambda}{(1+\lambda)^2} + \frac{(1+\lambda^2)}{2\lambda^2(1-\lambda)^2(\lambda^2-1)}\int_{\hat{l}}^{\hat{u}}(1+\lambda)y-3\lambda^2+\frac{4 \lambda^6}{(1+\lambda)^2}\frac{1}{y^2}\, dy = \nonumber \\
&\frac{2\lambda}{(1+\lambda)^2}+\frac{(1+\lambda^2)(\lambda-1)}{(\lambda^2-1)(\lambda+1)}=\frac{2\lambda}{(1+\lambda)^2}+\frac{1+\lambda^2}{(\lambda+1)^2}=1 \, .
\end{align}
We therefore conclude that
\begin{align}
\label{eindwinst}
P^{BNE}=\frac{1}{2}(d_+-s_-)\, .
\end{align}
When $\lambda = 1$, we have $A'(x,1)=\frac{3}{2}$, $l(1)=\frac{1}{4}$, $u(1)=\frac{3}{4}$ and $1-T(x,1)=\frac{1}{2}-2x$. Substituting these expressions in (\ref{totalBNE4}) again yields (\ref{eindwinst}).

\end{document}